\documentclass{article}%
\usepackage{amssymb}
\usepackage{amsmath}
\usepackage{amsfonts}
\usepackage{graphicx}%
\providecommand{\U}[1]{\protect\rule{.1in}{.1in}}
\newtheorem{theorem}{Theorem}
\newtheorem{acknowledgement}[theorem]{Acknowledgement}

\newtheorem{corollary}[theorem]{Corollary}

\newtheorem{definition}[theorem]{Definition}

\newtheorem{lemma}[theorem]{Lemma}
\newtheorem{notation}[theorem]{Notation}

\newtheorem{proposition}[theorem]{Proposition}
\newtheorem{remark}[theorem]{Remark}

\begin{document}

\title{The Hilbert $3/2$ Structure and Weil-Petersson Metric on the Space of the
Diffeomorphisms of the Circle Modulo Conformal Maps.}
\author{Maria Schonbek\\University of California\\Department of Mathematics\\Santa Cruz, CA 94109\\Institute of Mathematics,\\Bulgarian Academy of Sciences\\Sofia, Bulgaria
\and Andrey Todorov\thanks{The author was supported by Max-Plank Institute fur
Mathematik, Bonn during the preparation of this paper. }\\University of California\\Department of Mathematics\\Santa Cruz, CA 94109\\Institute of Mathematics,\\Bulgarian Academy of Sciences\\Sofia, Bulgaria
\and Jorge Zubelli\\IMPA\\Rio de Janeiro, Brazil}
\maketitle
\date{}

\begin{abstract}
The space of diffeomorphisms of the circle modulo boundary values of conformal
maps of the disk. The Sobolev $3/2$ norm on the tangent space space at $id$ of
the space of diffeomorphisms of the circle modulo boundary values of conformal
maps of the disk induces the only K\"{a}hler left invariant metric defined up
to a constant on the space of diffeomorphisms of the circle modulo boundary
values of conformal maps of the disk. We proved before that the completion of
the holomorphic tangent space at the identity of the space of diffeomorphisms
of the circle modulo boundary values of conformal maps of the disk with
respect to the Sobolev $3/2$ norm can be embedded as a closed Hilbert
susbspace into the tangent space at a point of the Segal-Wilson Grassmannian.
It was established in our previous paper that the natural metric on the
Segal-Wilson Grassmannian is K\"{a}hler has a negative sectional curvature in
the holomorphic direction bounded away from zero and the exponential map is a
complex analytic biholomorphic map from the tangent space of a given point of
the Segal-Wilson Grassmannian to the Segal-Wilson Grassmannian. In this paper
we prove that the completion of the space of diffeomorphisms of the circle
modulo boundary values of conformal maps of the disk with respect to the left
invariant K\"{a}hler metric is complex analytically isomorphic to the closed
totally geodesic closed Hilbert submanifold in the Segal-Wilson Grassmannian
using the exponential map. The new proof is much simpler and avoid the
technical proof that the change of the coordinates is complex analytic.

\end{abstract}
\tableofcontents

\section{Introduction}

\subsection{General Remarks}

The recent developments in string theory and the study of the theory of
infinite dimensional completely integrable systems renew the interest in the
study of differential geometric properties of infinite dimensional complex
manifolds. Donaldson in \cite{D} and \cite{D1} proposed the study of the
existence of canonical metrics on certain complex manifolds by the study of
the differential geometry of some infinite dimensional manifolds.

The geometric aspects of the space $\mathbf{T}^{\infty}:=\mathbf{Diff}%
_{+}^{\infty}\left(  S^{1}\right)  \left/  \mathbb{PSU}_{1,1}\right.  ,$ where
$\mathbf{Diff}_{+}^{\infty}\left(  S^{1}\right)  $ is the group of $C^{\infty
}$ diffeomorphisms of the circle preserving the orientations and,
$\mathbb{PSU}_{1,1}$ is the group of conformal maps of the unit disk
restricted to its boundary were studied in \cite{STZ}. It is a well known fact
that there exists a unique up to constant left invariant K\"{a}hler metric on
$\mathbf{T}^{\infty}.$ In \cite{STZ} we prove that the holomorphic curvature
of the left invariant K\"{a}hler metric on $\mathbf{T}^{\infty}$ is negative.
We prove that the completion of $\mathbf{T}^{\infty}$ with respect to the left
invariant K\"{a}hler metric defines Hilbert complex manifold structure on
$\mathbf{T}^{\infty}$ modeled by the Sobolev $3\left/  2\right.  $ norm.

In this paper we give a new proof of one of the main Theorems in \cite{STZ}:

\textbf{Theorem. }\textit{The completion }$\mathbf{T}^{3/2}$\textit{ of
}$\mathbf{T}^{\infty}$\textit{ with respect to the left invariant K\"{a}hler
metric defines Hilbert complex manifold structure on }$\mathbf{T}^{\infty}%
$\textit{ modeled by the Sobolev }$3\left/  2\right.  $\textit{ norm. }

The proof presented here is much simpler than the proof we gave in \cite{STZ}.
Our new proof is purely differential geometric. It is based on two facts. We
can embed the tangent space $T_{id,\mathbf{T}^{\infty}}$ at $id\in
\mathbf{T}^{\infty}$ into the tangent space at some point isomorphic to the
Hilbert space of Hilbert-Schmidt operators onto $\mathbf{T}^{\infty}$ of the
Segal-Wilson Grassmanian $\mathbb{G}\mathbf{r}_{\infty}$ isometrically. We
observed that the exponential map defined by the left invariant K\"{a}hler
metric on $\mathbb{G}\mathbf{r}_{\infty}$ is a complex analytic diffeomorphism
from the tangent space at any point onto $\mathbb{G}\mathbf{r}_{\infty}.$ Thus
$\mathbf{T}^{3/2}$ is the image of the completion of $T_{id,\mathbf{T}%
^{\infty}}$ in the tangent space at some point of $\mathbb{G}\mathbf{r}%
_{\infty}$ under the exponential map. Thus we avoid the technical check that
the transition functions of one open set to another are complex analytic functions.

In this paper we simplified also the proof of the embedding Theorem given in
\cite{STZ}. We gave a necessary and sufficient condition for the boundary
values on the unit circle of a holomorphic function to have $-3/2$ finite
Sobolev norm. This fact is simple and helps to simplify the proof of the
embedding Theorem.

A Theorem of Kuiper states that any Hilbert manifold is isomorphic to an open
set in a Hilbert space. We believe that our proof of the compatibilities of
the unique K\"{a}hler metric and the complex analytic Hilbert manifold modeled
after $3\left/  2\right.  $ Sobolev, norm is very natural and simple in view
of the above Theorem of Kuiper.

Some of the results in the paper \cite{TT} are closely related to our results
obtained in \cite{STZ}.

We have included an Appendix in which we gave proofs of the two main results
obtained in \cite{STZ} about the properties of the left invariant metric that
we use in the paper. This is done to make the paper completely self contained.

\begin{acknowledgement}
The second author would like to thank R. Friedrich and B. Khesin for their
interest in this work and the help in improving the paper. R. Friedrich
pointed out that the paper \cite{M} maybe is related to our paper.
\end{acknowledgement}

\subsection{Outline of the Proof of the Existence of Sobolev $3/2$ Hilbert
Complex Structure on $\mathbf{T}^{\infty}$}

The new proof of that the completion of $\mathbf{T}^{\infty}$ with respect to
the left invariant K\"{a}hler metric defines Hilbert complex manifold
structure on $\mathbf{T}^{\infty}$ modeled by the Sobolev $3\left/  2\right.
$ norm is based on three facts:

\textbf{Fact 1.} Let $\mu_{\psi}(z):=\left(  1-\left\vert z\right\vert
^{2}\right)  ^{2}\psi\left(  \overline{z}\right)  ,$ $\psi\left(  \overline
{z}\right)  $ is an antiholomorphic function on the unit disk $\mathbb{D},$
and
\begin{equation}
\left.  \psi\left(  \overline{z}\right)  \right\vert _{S^{1}}\in
\mathbf{H}_{S^{1},h}^{-3/2}, \label{1A}%
\end{equation}
where $\mathbf{H}_{S^{1},h}^{-3/2}$ is the space of the restrictions of
holomorphic functions on the unit circle with finite $-3/2$ Sobolev norm. Then
we have%
\begin{equation}
\underset{|z|<1}{\sup}\left\vert \mu_{\psi}(z)\right\vert =\left\Vert
\psi\left(  \overline{z}\right)  \right\Vert _{\mathbf{L}^{\infty}%
(\mathbb{D})}<C_{\psi}<\infty. \label{1a}%
\end{equation}
This fact was proved also independently and very elegantly in \cite{TT}.

\textbf{Fact 2. }Let us define%
\[
\mathbb{H}^{-1,1}\left(  \mathbb{D}\right)  :=
\]%
\begin{equation}
\left\{  \mu_{\psi}(z)\left\vert \left\Vert \left\vert \mu_{\psi
}(z)\right\vert \right\Vert _{\mathbf{L}_{\mathbb{D}}^{\infty}}<\infty,\text{
}\partial\psi\left(  \overline{z}\right)  =0,\text{ }\ \text{\& }\left.
\psi\left(  \overline{z}\right)  \right\vert _{S^{1}}\in\mathbf{H}_{S^{1}%
}^{-3/2}\right.  \right\}  . \label{2a}%
\end{equation}
We will define a Hilbert structure on $\mathbb{H}^{-1,1}\left(  \mathbb{D}%
\right)  .$ It will be based on the following non degenerate $\mathbf{L}^{2}$
pairing on the unit disk $\mathbb{D},$ $\mathbb{H}^{-1,1}\left(
\mathbb{D}\right)  \times\mathbf{H}_{S^{1},h}^{-3/2}\rightarrow\mathbb{C}$
given by%
\[
\left.  \left\langle \mu_{\psi}(z),\eta\left(  \overline{z}\right)
\right\rangle \right\vert _{S^{1}}=\frac{\sqrt{-1}}{2\pi}%
{\displaystyle\int\limits_{\mathbb{D}}}
\left(  1-\left\vert z\right\vert ^{2}\right)  ^{2}\psi\left(  \overline
{z}\right)  (z)\overline{\eta\left(  \overline{z}\right)  }dz\wedge
\overline{dz}=
\]%
\begin{equation}
\frac{\sqrt{-1}}{2\pi}%
{\displaystyle\int\limits_{\mathbb{D}}}
\mu_{\psi}(z)\overline{\eta\left(  z\right)  }dz\wedge\overline{dz}.
\label{6A}%
\end{equation}
Direct computations show
\[
\left\vert \frac{\sqrt{-1}}{2\pi}%
{\displaystyle\int\limits_{\mathbb{D}}}
\mu_{\psi}(z)\overline{\eta\left(  z\right)  }dz\wedge\overline{dz}\right\vert
=\left\vert \left\langle \left.  \psi\left(  \overline{z}\right)  \right\vert
_{S^{1}},\left.  \eta\left(  \overline{z}\right)  \right\vert _{S^{1}%
}\right\rangle _{\mathbf{H}_{S^{1},h}^{-3/2}}\right\vert <\infty.
\]

$\mathbb{H}^{-1,1}\left(  \mathbb{D}\right)  $ is the $\mathbf{L}^{2}$ dual to
$\mathbf{H}_{S^{1},h}^{-3/2}$ by the pairing $\left(  \ref{6A}\right)  $ and
thus it is isomorphic to $\mathbf{H}_{S^{1},h}^{3/2}.$ We will give an
explicit isomorphism between $\mathbb{H}^{-1,1}\left(  \mathbb{D}\right)  $
and $\mathbf{H}_{S^{1},h}^{3/2}.$ Let
\begin{equation}
\psi\left(  \overline{z}\right)  =%
{\displaystyle\sum\limits_{n\geq2}}
a_{n}\overline{z}^{n-2}\text{ and }\Psi\left(  \frac{1}{z}\right)  =%
{\displaystyle\sum\limits_{n\geq2}}
\frac{a_{n}}{n(n+1)(n+2)}\left(  \frac{1}{z}\right)  ^{n-2}\text{ .}
\label{3a}%
\end{equation}
It is easy to see that $\left(  \ref{1A}\right)  $ implies
\begin{equation}
\left.  \Psi\left(  \frac{1}{z}\right)  \right\vert _{S^{1}}\in\mathbf{H}%
_{S^{1}}^{3/2} \label{4a}%
\end{equation}
Consider the map $\digamma:\mathbb{H}^{-1,1}\left(  \mathbb{D}\right)
\rightarrow\mathbf{H}_{S^{1}}^{3/2}$ given by
\begin{equation}
\digamma\left(  \mu_{\psi}(z)\right)  =\left.  \Psi\left(  \frac{1}{z}\right)
\right\vert _{S^{1}}. \label{5a}%
\end{equation}
We define a Hilbert norm of $\mathbb{H}^{-1,1}\left(  \mathbb{D}\right)  $ as
follows:
\begin{equation}
\left\Vert \mu_{\psi}(z)\right\Vert _{\mathbb{H}^{-1,1}\left(  \mathbb{D}%
\right)  }^{2}=\left\Vert \Psi(z)\right\Vert _{\mathbf{H}_{S^{1}}^{3/2}}^{2}.
\label{6a}%
\end{equation}
We will show that completion of the tangent space $T_{id,\mathbf{T}^{\infty}}$
with respect to the left invariant metric is isomorphic to the Hilbert space
$\mathbb{H}^{-1,1}\left(  \mathbb{D}\right)  $ with the norm defined by
$\left(  \ref{6a}\right)  .$

This observation has the following interpretation; The space $\mathbf{H}%
_{S^{1},h}^{-3/2}$ can be identified with a quadratic differentials%
\[
\psi\left(  \frac{1}{z}\right)  \left(  d\left(  \frac{1}{z}\right)  \right)
^{\otimes2}%
\]
which is isomorphic to completion of the holomorphic cotangent space
$\Omega_{\mathbf{T}^{\infty}}^{1,0}$ with respect to the left invariant
K\"{a}hler metric when $\left.  \psi\left(  \frac{1}{z}\right)  \right\vert
_{S^{1}}\in\mathbf{H}_{S^{1},h}^{-3/2}.$ $\mathbb{H}^{-1,1}\left(
\mathbb{D}\right)  \approxeq\mathbf{H}_{S^{1},h}^{3/2}$ can be identified with
the completion of the tangent space $T_{id,\mathbf{T}^{\infty}}$ and the
duality is given by the contraction of $\psi\left(  \frac{1}{z}\right)
\left(  d\left(  \frac{1}{z}\right)  \right)  ^{\otimes2}$ with the Poincare
metric on $\mathbb{D}.$ The Poincare metric induces the left invariant
K\"{a}hler metric on $\mathbf{T}^{\infty}.$ Thus the space $\mathbb{H}%
^{-1,1}\left(  \mathbb{D}\right)  $ consists of harmonic Beltrami
differentials $\mu_{\psi}\left(  \overline{dz}\otimes\frac{d}{dz}\right)  $
with respect to left invariant K\"{a}hler metric.

\textbf{Fact 3. }In this paper we give a new and simpler proof of the
following fact proved in \cite{STZ}; The map
\[
\iota:\mu_{\psi}(z)\rightarrow\ker\left(  \overline{\partial}-\mu_{\psi
}\partial\right)
\]
where $\mu_{\psi}$ is such that $\left\Vert \mu_{\psi}\right\Vert
_{\mathbf{L}^{\infty}\left(  \mathbb{D}\right)  }\leq k<1$ is an embedding
into $T_{\mathbf{H}^{+},\mathbb{G}r_{\infty}}.$ From here we can conclude that
$\iota\left(  \mathbb{H}^{-1,1}\left(  \mathbb{D}\right)  \right)  $ is closed
Hilbert subspace in the tangent space $T_{\mathbf{H}^{+},\mathbb{G}r_{\infty}%
}$ which is the Hilbert space of Hilbert-Scmidt operators $T:\mathbf{H}%
^{+}\rightarrow\mathbf{H}^{-}.$ $\mathbf{H}^{+}(\mathbf{H}^{-})$ are the space
of the boundary values holomorphic (antiholomorphic) functions in the unit
disk with a finite $\mathbf{L}^{2}(S^{1})$ norm. From here we can conclude
that $\iota\left(  \mathbb{H}^{-1,1}\left(  \mathbb{D}\right)  \right)  $ is
closed Hilbert subspace in the tangent space $T_{\mathbf{H}^{+},\mathbb{G}%
r_{\infty}}.$

We observed in \cite{STZ} that the left invariant metric on $\mathbb{G}%
r_{\infty}$ is K\"{a}hler and it has a negative holomorphic sectional
curvature bounded away from zero. From this fact we derived that the
exponential map
\[
\exp:T_{\mathbf{H}^{+},\mathbb{G}r_{\infty}}\rightarrow\mathbb{G}r_{\infty}%
\]
is a complex analytic diffeomorphism.

Nag proved that $\mathbf{T}^{\infty}$ together with the left invariant
K\"{a}hler metric can be embedded isometrically into $\mathbb{G}r_{\infty}.$
Thus we can conclude that the left invariant K\"{a}hler metric on
$\mathbf{T}^{\infty}$ modeled by the Sobolev $3/2$ space of $T_{id,\mathbf{T}%
^{\infty}}$ has a negative holomorphic sectional curvature bounded away from
zero. The result of Nag implies that the completion $\mathbf{T}^{3/2}$ of
$\mathbf{T}^{\infty}$ with respect to the left invariant K\"{a}hler metric on
$\mathbf{T}^{\infty}$ will be complex analytically isomorphic to the totally
geodesic complex analytic Hilbert submanifold $\exp\left(  \iota\left(
\mathbb{H}^{-1,1}\left(  \mathbb{D}\right)  \right)  \right)  $ in
$\mathbb{G}r_{\infty}.$

\section{Preliminary Material}

\subsection{Basic Notations and Definitions}

\begin{notation}
\label{0}Let $\mathbb{D}:=\left\{  z\in\mathbb{C}\left\vert \left\vert
z\right\vert <1\right.  \right\}  $ and $\mathbb{D}^{\ast}:=\left\{  \frac
{1}{z}\in\mathbb{C}\left\vert \left\vert z\right\vert <1\right.  \right\}  .$
We recall that the Schwarzian derivative of an analytic function $f$ is
defined by
\[
\left[  \mathcal{S}(f)\right]  (z)\overset{def}{=}\frac{f^{\prime\prime\prime
}}{f^{\prime}}-\frac{3}{2}\left(  \frac{f^{\prime\prime}}{f^{\prime}}\right)
^{2}.
\]

\end{notation}

\begin{definition}
\label{B1}Let $\psi(\overline{z})$ be antiholomorphic function in the unit
disk. Suppose that $\underset{z\in\mathbb{D}}{\sup}\left\vert \left(
1-\left\vert z\right\vert ^{2}\right)  ^{2}\psi(\overline{z})\right\vert \leq
k<1.$ Define
\begin{subequations}
\begin{equation}
\mu_{\psi}(z)=\left\{
\begin{array}
[c]{c}%
\left(  1-\left\vert z\right\vert ^{2}\right)  ^{2}\psi(\overline{z}),\text{
}z\in\mathbb{D}\\
0,\text{ \ \ \ \ \ }z\notin\mathbb{D}%
\end{array}
\right.  . \label{bel1}%
\end{equation}
Then by abuse of notation define the Beltrami differential
\end{subequations}
\[
\mu_{\psi}:=\mu_{\psi}(z)\left(  \overline{dz}\otimes\frac{\partial}{\partial
z}\right)  .
\]

\end{definition}

\begin{notation}
We will call the operator $\overline{\partial}_{\mu_{\psi}}:=\overline
{\partial}-\mu_{\psi}(z)\partial$ Beltrami operator. This space we will denote
it by $\mathbb{H}_{1}^{1,1}\left(  \mathbb{D}\right)  .$
\end{notation}

We will used the following Theorem of Ahlfors and Weill stated bellow. The
proof can be found in page 100 of \cite{gardiner}:

\begin{theorem}
\label{AW} Let $\Psi\left(  z\right)  :\mathbb{D}^{\ast}\rightarrow
\overline{\mathbb{C}}$ be an univalent function that can be extended to a
quasi-conformal map of the Riemann sphere $\mathbb{CP}^{1}.$ Let
\begin{equation}
\Psi\left(  z\right)  =z+%
{\displaystyle\sum\limits_{k=1}^{\infty}}
a_{k}\left(  \frac{1}{z}\right)  ^{k},\text{ }\mathcal{S}\left[  \left.
\Phi\left(  z\right)  \right\vert _{\mathbb{D}^{\ast}}\right]  =%
{\displaystyle\sum\limits_{k\geq4}^{\infty}}
b_{k}\left(  z\right)  ^{-k} \label{aw0}%
\end{equation}
for $|z|<1.$ Let%
\begin{equation}
\mu_{\psi}(z)=\left\{
\begin{array}
[c]{cc}%
-\frac{1}{2}\left(  1-|z|^{2}\right)  ^{2}\frac{\mathcal{S}\left[  \Phi\left(
z\right)  \left\vert _{\mathbb{D}^{\ast}}\right.  \right]  \left(
\overline{z}\right)  }{z^{4}}, & \text{ }|z|<1\\
\text{ \ \ \ }0,\text{ \ \ \ \ \ \ \ \ \ \ \ \ \ \ \ \ \ \ \ \ \ \ \ } &
\ \text{ }|z|\geq1
\end{array}
\right.  \text{\ \ \ \ \ \ }\ \ \ \ \text{\ \ \ \ \ } \label{aw}%
\end{equation}
Then, for the quasi-conformal mapping $\Phi_{\mu_{\psi}}$ on the Riemann
sphere satisfying%
\[
\left(  \overline{\partial}-\mu_{\psi}(z)\partial\right)  \Phi_{\mu_{\psi}%
}(z)=0
\]
we have that $\psi\left(  z\right)  =\mathcal{S}\left[  \left.  \Phi\left(
z\right)  \right\vert _{\mathbb{D}^{\ast}}\right]  $ and $\left.  \Phi
_{\mu_{\psi}}\left(  z\right)  \right\vert _{\mathbb{D}^{\ast}}=\Psi\left(
z\right)  $ is the univalent function on $\mathbb{D}^{\ast}$ defined by
$\left(  \ref{aw0}\right)  $.
\end{theorem}

\begin{definition}
We will define: $\mathbf{T}^{\infty}:=\mathbf{Diff}_{+}^{\infty}\left(
S^{1}\right)  \left/  \mathbb{PSU}_{1,1}\right.  .$
\end{definition}

\begin{definition}
\label{com str}Let M be an even dimensional C$^{\infty}$ manifold, $T$ be the
tangent bundle \ and $T^{\ast}$ be the cotangent bundle on M. We will say that
M has an almost complex structure if there exists a section $I\in C^{\infty
}(M,Hom(T^{\ast},T^{\ast}))$ such that $I^{2}=-id.$
\end{definition}

This definition is equivalent to the following one:

\begin{definition}
Let M be an even-dimensional C$^{\infty}$ manifold. If $T_{\text{M}}^{\ast
}\otimes\mathbb{C}=\Omega_{\text{M}}^{1,0}\oplus\overline{\Omega_{\text{M}%
}^{1,0}}$ is a global splitting of the complexified cotangent bundle such that
$\Omega_{\text{M}}^{0,1}=\overline{\Omega_{\text{M}}^{1,0}}$, then we will say
that M has an almost complex structure.
\end{definition}

\begin{definition}
\label{ICS}We will say that an almost complex structure is an integrable one,
if for each point $x\in$M there exists an open set $U\subset$M such that we
can find local coordinates $z^{1},..,z^{n}$ such that $dz^{1},..,dz^{n}$ \ are
linearly independent at each point $m\in U$ and they generate $\Omega
^{1,0}|_{U}.$
\end{definition}

\begin{notation}
\label{B0}We define:
\begin{equation}
\mathbf{H}^{+}=\left\{  f(z)\in H(\mathbb{D})|\text{ }f(z)=%
{\displaystyle\sum\limits_{n\geq0}}
a_{n}z^{n}\text{ \& }\left\Vert f(z)\right\Vert ^{2}=%
{\displaystyle\sum\limits_{n\geq0}}
\left\vert a_{n}\right\vert ^{2}<\infty\right\}  \label{h+}%
\end{equation}
and
\begin{equation}
\mathbf{H}^{-}=\left\{  g(z)\in H(\mathbb{D}^{\ast})|\text{ }f\left(
z\right)  =%
{\displaystyle\sum\limits_{n\geq1}}
b_{n}z^{-n}\text{ \& }\left\Vert g\left(  z\right)  \right\Vert ^{2}=%
{\displaystyle\sum\limits_{n\geq0}}
\left\vert b_{n}\right\vert ^{2}<\infty\right\}  . \label{h-}%
\end{equation}

\end{notation}

\begin{definition}
\label{1}We define $\mathbb{H}_{S^{1},h}^{\alpha}$ to be the space of all
holomorphic functions $g(z)$ on $\mathbb{D}^{\ast}$ such that the restriction
of $g\left(  z\right)  $ on the boundary $S^{1}$ of $\mathbb{D}^{\ast}$ is an
element of the Hilbert space with a Sobolev $\alpha$ norm, i.e.%
\[
\left\Vert g\right\Vert _{\alpha}^{2}=%
{\displaystyle\sum\limits_{n=0}^{\infty}}
n^{2\alpha}\left\vert b_{n}\right\vert ^{2}<\infty,
\]
where $\alpha$ is any real number.
\end{definition}

\begin{remark}
In this paper, for convenience, we shall use the following notation to the
norms $3/2$ and $-3/2$, which are equivalent to the Sobolev $3/2$ and $-3/2$
norms:
\[
\left\Vert g\right\Vert _{3/2}^{2}:=%
{\displaystyle\sum\limits_{n=2}^{\infty}}
n\left(  n^{2}-1\right)  \left\vert a_{n}\right\vert ^{2}\text{ and
}\left\Vert g\right\Vert _{-3/2}^{2}:=%
{\displaystyle\sum\limits_{n=2}^{\infty}}
\frac{\left\vert a_{n}\right\vert ^{2}}{n\left(  n^{2}-1\right)  }.
\]

\end{remark}

\begin{remark}
\label{2}Let $\psi_{i}(\theta)=%
{\displaystyle\sum\limits_{k\geq2}^{\infty}}
a_{k}^{i}e^{-i(k-2)\theta}\in\mathbb{H}_{S^{1},h}^{p\left/  q\right.  }$ be
two complex valued functions on $S^{1}.$ The $\mathbf{L}_{S^{1}}^{2}$ pairing
between $\mathbb{H}_{S^{1},h}^{p\left/  q\right.  }$ and $\mathbb{H}_{S^{1}%
,h}^{-p\left/  q\right.  }$ is defined by
\begin{equation}
\left\langle \psi_{1},\psi_{2}\right\rangle =%
{\displaystyle\sum\limits_{k=2}^{\infty}}
a_{k}^{1}\overline{a_{k}^{2}}. \label{l2}%
\end{equation}
The pairing $\left(  \ref{l2}\right)  ~$is a non-degenerate pairing.
\end{remark}

\subsection{Left Invariant K\"{a}hler Structures on $\mathbf{Diff}_{+}%
^{\infty}(S^{1})\left/  \mathbb{PSU}_{1,1}\right.  $}

The complexified tangent space $T_{id,\mathbf{Diff}_{+}^{\infty}(S^{1})\left/
\mathbb{PSU}_{1,1}\right.  }$ at the identity can be described as follows:
\[
T_{id,\mathbf{Diff}_{+}^{\infty}(S^{1})\left/  \mathbb{PSU}_{1,1}\right.
}\otimes\mathbb{C}=\left\{  f(\theta)\frac{d}{d\theta}\left\vert f(\theta)=%
{\displaystyle\sum\limits_{n\neq0,\pm1}}
a_{n}e^{in\theta}\right.  \right\}  .
\]
Thus we have a splitting:%
\[
T_{id,\mathbf{Diff}_{+}^{\infty}(S^{1})\left/  \mathbb{PSU}_{1,1}\right.
}\otimes\mathbb{C}=T_{id,\mathbf{Diff}_{+}^{\infty}(S^{1})\left/
\mathbb{PSU}_{1,1}\right.  }^{1,0}\oplus\overline{T_{id,\mathbf{Diff}%
_{+}^{\infty}(S^{1})\left/  \mathbb{PSU}_{1,1}\right.  }^{1,0}},
\]
where%
\[
T_{id,\mathbf{Diff}_{+}^{\infty}(S^{1})\left/  \mathbb{PSU}_{1,1}\right.
}^{1,0}=\left\{  f(\theta)\frac{d}{d\theta}\left\vert f(\theta)=%
{\displaystyle\sum\limits_{n>1}}
c_{n}e^{i\left(  n-2\right)  \theta}\right.  \right\}  .
\]
The complexified tangent space $T_{id,\mathbf{Diff}_{+}^{\infty}(S^{1})\left/
\mathbb{PSU}_{1,1}\right.  }$ at the identity is presented as follows
\[
T_{id,\mathbf{Diff}_{+}^{\infty}(S^{1})\left/  \mathbb{PSU}_{1,1}\right.
}=T_{id,\mathbf{Diff}_{+}^{\infty}(S^{1})\left/  \mathbb{PSU}_{1,1}\right.
}^{1,0}\oplus\overline{T_{id,\mathbf{Diff}_{+}^{\infty}(S^{1})\left/
\mathbb{PSU}_{1,1}\right.  }^{1,0}}.
\]
Thus a left invariant almost complex structure on the coadjoint orbit and
$\mathbf{T}^{\infty}:=\mathbf{Diff}_{+}^{\infty}(S^{1})\left/  \mathbb{PSU}%
_{1,1}\right.  $ was defined. It is proved in \cite{nagver} that this complex
structure is integrable.

We will define the analogue of the Weil-Petersson metric on $\mathbf{T}%
^{\infty}.$ It is proved in \cite{kirillov}, \cite{M}$,$ \cite{witten1} and
\cite{witten2} that the Weil-Petersson metric is unique left invariant
K\"{a}hler metric defined up to a constant on $\mathbf{T}^{\infty}.$

\begin{definition}
Let
\[
f(\theta)\frac{d}{d\theta}=\left(
{\displaystyle\sum\limits_{n>1}}
a_{n}e^{i\left(  n-2\right)  \theta}\right)  \frac{d}{d\theta}\in
T_{id,\mathbf{Diff}_{+}^{\infty}(S^{1})\left/  \mathbb{PSU}_{1,1}\right.
}^{1,0}.
\]
Then the norm$\left\Vert f(\theta)\frac{d}{d\theta}\right\Vert _{W.P.}$ of
$f(\theta)\frac{d}{d\theta}$ is given by%
\begin{equation}
\left\Vert f(\theta)\frac{d}{d\theta}\right\Vert _{W.P.}^{2}=%
{\displaystyle\sum\limits_{n>1}}
\frac{n\left(  n^{2}-1\right)  }{12}\left\vert a_{n}\right\vert ^{2}.
\label{3/2pair}%
\end{equation}

\end{definition}

\section{Preliminary Results}

\subsection{$\mathbf{H}_{S^{1},h}^{-3\left/  2\right.  }$ and the
$\mathbf{L}^{\infty}$ Norm}

\begin{theorem}
\label{SS0}Let $\mathbf{H}_{S^{1},h}^{-3\left/  2\right.  }$ be the Hilbert
space defined by Definition \ref{1}. Let $|z\}\leq1.$ Suppose that
$\psi\left(  z\right)  =%
{\displaystyle\sum\limits_{n=2}^{\infty}}
a_{n}\left(  \frac{1}{z}\right)  ^{n-2}$ is a holomorphic function defined on
$\mathbb{D}^{\ast}$ such that $\psi\left(  z\right)  \left\vert _{S^{1}%
}\right.  \in\mathbf{H}_{S^{1},h}^{-3\left/  2\right.  }.$ Let $\psi\left(
\overline{z}\right)  =%
{\displaystyle\sum\limits_{n=2}^{\infty}}
a_{n}\left(  \overline{z}\right)  ^{n-2}.$ Then $\left(  1-|z|^{2}\right)
^{2}|\psi\left(  \overline{z}\right)  |<C$ for $z\in\mathbb{D},$ i.e. $|z|<1.$
\end{theorem}

\textbf{Proof: }The proof of Theorem \ref{SS0} can be find in \cite{TT}. We
will repeat the proof given in \cite{TT}. The condition $\psi(z)\left\vert
_{S_{1}}\right.  \in\mathbf{H}_{S^{1},h}^{-3\left/  2\right.  }\ $\ means that
we have
\begin{equation}
\left\Vert \psi\right\Vert _{\mathbf{H}_{S^{1},h}^{-3\left/  2\right.  }}%
^{2}:=%
{\displaystyle\sum\limits_{n\geq2}}
\frac{\left\vert a_{n}\right\vert ^{2}}{n\left(  n^{2}-1\right)  }%
<\infty\text{ .} \label{H3/2}%
\end{equation}
Let $b_{n}:=\frac{a_{n}}{\sqrt{n\left(  n^{2}-1\right)  }}.$ Then
Cauchy-Schwarz inequality implies that
\begin{equation}
\left\vert \psi\left(  \overline{z}\right)  \right\vert \leq\left\vert
{\displaystyle\sum\limits_{n=2}^{\infty}}
n\left(  n^{2}-1\right)  \left\vert b_{n}\right\vert ^{2}\right\vert
^{1\left/  2\right.  }\left\vert
{\displaystyle\sum\limits_{n=2}^{\infty}}
n\left(  n^{2}-1\right)  \left\vert \overline{z}\right\vert ^{2n-4}\right\vert
^{1\left/  2\right.  } \label{h2}%
\end{equation}
for every $z\in\mathbb{D}.$ Since%
\begin{equation}
\left\vert
{\displaystyle\sum\limits_{n=2}^{\infty}}
n\left(  n^{2}-1\right)  \left\vert \overline{z}\right\vert ^{2n-4}\right\vert
=\frac{6}{\left(  1-\left\vert z\right\vert ^{2}\right)  ^{2}} \label{h3}%
\end{equation}
we get that $\left(  \ref{h2}\right)  $ and $\left(  \ref{h3}\right)  $ imply
that $\left(  1-|z|^{2}\right)  ^{2}\left\vert \psi\left(  \overline
{z}\right)  \right\vert \leq C.$ So Theorem \ref{SS0} is proved.
$\blacksquare$

\begin{theorem}
\label{SS1}Suppose that $\psi\left(  \overline{z}\right)  =%
{\displaystyle\sum\limits_{n=2}^{\infty}}
a_{n}\overline{z}^{n}$ is a holomorphic function defined on $\mathbb{D}$ such
that $\psi\left(  \overline{z}\right)  \left\vert _{S^{1}}\right.
\in\mathbf{H}_{S^{1},h}^{-3\left/  2\right.  }.$ Then for some real $\alpha$
such that $0<\alpha<1$ we have
\begin{equation}
\underset{\underset{|z|<1}{|z|\rightarrow1}}{\lim}\frac{\left\vert \psi\left(
\left\vert \overline{z}\right\vert \right)  \right\vert }{\left(  1-\left\vert
z\right\vert \right)  ^{a}}=C_{\psi}<\infty\label{h4}%
\end{equation}

\end{theorem}

\textbf{Proof: }Since $\psi\left(  \overline{z}\right)  $ is an atiholomorphic
function, then the maximum principle shows that the function $\underset
{r=|z|<1}{\max}\left\vert \psi\left(  \left\vert z\right\vert \right)
\right\vert =\psi(r)$ is an increasing function. Suppose that $\underset
{r\rightarrow\infty}{\lim}\psi(r)=\infty.$ Theorem \ref{SS} implies that
\[
\left(  1-\left\vert r\right\vert ^{2}\right)  ^{2}\psi(r)=\left(
1-\left\vert r\right\vert \right)  ^{2}\left(  1+\left\vert r\right\vert
\right)  ^{2}\psi(r)<C.
\]
So we get that
\begin{subequations}
\begin{equation}
\underset{r\rightarrow\infty}{\lim}\frac{\psi(r)}{\left(  1-\left\vert
r\right\vert \right)  ^{\alpha}}=C_{\psi}<\infty\label{abv}%
\end{equation}
for some $0<\alpha<2.$ Easy computations in polar coordinates show that%
\end{subequations}
\[%
{\displaystyle\int\limits_{\mathbb{D}}}
\left(  1-\left\vert z\right\vert ^{2}\right)  ^{2}\left\vert \psi
(\overline{z})\right\vert ^{2}dz\wedge\overline{dz}=\left\Vert \psi\left(
\overline{z}\right)  \left\vert _{S^{1}}\right.  \right\Vert _{\mathbf{H}%
_{S^{1},h}^{-3\left/  2\right.  }}^{2}.
\]
Thus the assumption that $\psi\left(  \overline{z}\right)  \left\vert _{S^{1}%
}\right.  \in$ $\mathbf{H}_{S^{1},h}^{-3\left/  2\right.  }$ implies that%
\begin{equation}
\left\vert
{\displaystyle\int\limits_{\mathbb{D}}}
\left(  1-\left\vert z\right\vert ^{2}\right)  ^{2}\left\vert \psi
(\overline{z})\right\vert ^{2}dz\wedge\overline{dz}\right\vert <\infty
\label{abv1}%
\end{equation}
Using polar coordinates $\left(  \ref{abv1}\right)  $ implies that
\[%
{\displaystyle\int\limits_{0}^{1}}
\left(  \left(  1-\left\vert r\right\vert \right)  \left(  1+\left\vert
r\right\vert \right)  \left\vert \psi(r)\right\vert \right)  dr<4%
{\displaystyle\int\limits_{0}^{1}}
\left(  \left(  1-\left\vert r\right\vert \right)  \left\vert \psi
(r)\right\vert \right)  dr<\infty.
\]
The last inequality and $\left(  \ref{abv}\right)  $ imply that
\[
\underset{0<r<1,\text{ }r\rightarrow1}{\lim}\underset{|z|=|r|}{\sup}\left\vert
\psi(z)\right\vert \left(  1-r\right)  ^{-\alpha}=C>0
\]
for $0<\alpha<1.$ Theorem \ref{SS1} is proved$.$ $\blacksquare$

\begin{corollary}
Let $\psi(\overline{z})\left\vert _{S^{1}}\right.  \in\mathbf{H}_{S^{1}%
,h}^{-3\left/  2\right.  }.$ Then
\begin{equation}
\left.  \left(  1-|z|^{2}\right)  ^{2}\psi(\overline{z})\left(  \frac{d}%
{dz}\otimes\overline{dz}\right)  \right\vert _{\mathbb{D}^{\ast}}=0.
\label{cs3a}%
\end{equation}

\end{corollary}

\subsection{Duality}

\begin{theorem}
\label{Pair} \textit{Let }$\psi(\overline{z})=%
{\displaystyle\sum\limits_{n=2}^{\infty}}
a_{n}\left(  \overline{z}\right)  ^{n-2}$ and $\psi(\overline{z})\left\vert
_{S^{1}}\right.  \in\mathbf{H}_{S^{1},h}^{-3\left/  2\right.  }.$ \textit{Let
us define the linear functional} $L_{\psi}$ \textit{on }$\mathbf{H}_{S^{1}%
,h}^{-3\left/  2\right.  }$ \textit{as follows:}
\begin{equation}
L_{\psi}(\eta(z)):=\frac{1}{2\pi i}%
{\displaystyle\int\limits_{\mathbb{D}}}
\left(  1-\left\vert z\right\vert ^{2}\right)  ^{2}\psi(\overline{z}%
)\overline{\eta\left(  \overline{z}\right)  }dz\wedge\overline{dz}. \label{lf}%
\end{equation}
\textit{where }$\eta(z)$ is a holomorphic function on $\mathbb{D}^{\ast}$ such
that $\eta(z)\left\vert _{S^{1}}\right.  \in\mathbf{H}_{S^{1},h}^{-3\left/
2\right.  }.$ \textit{Then }
\end{theorem}

\textit{\textbf{1. }The linear functional }$L_{\psi}(\eta(z))$\textit{ defined
by }$\left(  \ref{lf}\right)  $ \textit{is continuous, }

\textbf{2. }\textit{The map }%
\[
\psi(z)\left\vert _{S^{1}}\right.  =\psi(\overline{z})\left\vert _{S^{1}%
}\right.  \in\mathbf{H}_{S^{1},h}^{-3\left/  2\right.  }\rightarrow L_{\psi
}\in\mathbf{H}_{S^{1},}^{3\left/  2\right.  }=Hom\left(  \mathbf{H}_{S^{1}%
,h}^{-3\left/  2\right.  },\mathbb{C}\right)
\]
\textit{is a canonical isomorphism between }$\mathbf{H}_{S^{1},h}^{-3\left/
2\right.  }$ \textit{and its dual} $\mathbf{H}_{S^{1},h}^{3\left/  2\right.
}$ with respect to \textit{the }$\mathbf{L}_{S^{1}}^{2}.$

\textit{\textbf{3}. Let }$\Psi\left(  \overline{z}\right)  $\textit{ be
defined by }$\left(  \ref{3a}\right)  .$\textit{ Then we have the following
formula:}%
\begin{equation}
L_{\psi}(\eta(z))=\left\langle \Psi\left(  \theta\right)  ,\eta(\theta
)\right\rangle _{\mathbf{L}^{2}\left(  S^{1}\right)  }=\left\langle
\psi(\theta),\eta(\theta)\right\rangle _{\mathbf{H}_{S^{1},h}^{-3\left/
2\right.  }}. \label{l2da}%
\end{equation}
\textit{The map }$\psi\rightarrow\Psi$ \textit{defines an isomorphism between
}$\mathbf{H}_{S^{1},h}^{3\left/  2\right.  }$ \textit{and }$Hom\left(
\mathbf{H}_{S^{1},h}^{-3\left/  2\right.  },\mathbb{C}\right)  .$

\textbf{Proof:} Now we will proceed with proof of Theorem \ref{Pair}.

\begin{lemma}
\label{Pair0}Let $|z|<1.$ Let $\psi\left(  z\right)  =%
{\displaystyle\sum\limits_{n>1}}
a_{n}\left(  \frac{1}{z}\right)  ^{n-2}$ be a holomorphic function in
$\mathbb{D}^{\ast}$ such that $\psi\left(  z\right)  \left\vert _{S^{1}%
}\right.  \in\mathbf{H}_{S^{1},h}^{-3\left/  2\right.  }.$ Let us define
$\Psi\left(  \overline{z}\right)  $ by $\left(  \ref{3a}\right)  .$ Then
\[
\Psi\left(  \overline{z}\right)  \left\vert _{S^{1}}\right.  =\Psi\left(
\theta\right)  \in\mathbf{H}_{S^{1},h}^{3\left/  2\right.  }\text{ }\&\text{
}\left\Vert \Psi\right\Vert _{\mathbf{H}_{S^{1},h}^{3\left/  2\right.  }}%
^{2}=\left\Vert \psi\right\Vert _{\mathbf{H}_{S^{1},h}^{-3\left/  2\right.  }%
}^{2}.
\]

\end{lemma}

\textbf{Proof: }Let $\psi\left(  \overline{z}\right)  =%
{\displaystyle\sum\limits_{n\geq2}^{\infty}}
a_{n}(\overline{z})^{n-2}$ be the antiholomorphic function in the unit disk
$\mathbb{D}$ defined by the coefficients $a_{n}$ of $\psi(z)$ by $.$ The
assumption $\psi\left(  z\right)  \left\vert _{S^{1}}\right.  \in
\mathbf{H}_{S^{1},h}^{-3\left/  2\right.  }$ implies that
\[
\left\Vert \psi\right\Vert _{\mathbf{H}_{S^{1},h}^{-3\left/  2\right.  }}^{2}=%
{\displaystyle\sum\limits_{n=2}^{\infty}}
\frac{\left\vert a_{n}\right\vert ^{2}}{n\left(  n^{2}-1\right)  }<\infty.
\]
The definition $\left(  \ref{3a}\right)  $ of $\Psi\left(  \overline
{z}\right)  =%
{\displaystyle\sum\limits_{n=0}^{\infty}}
b_{n}\left(  \overline{z}\right)  ^{n}$ states that the coefficients $b_{n}$
of $\Psi\left(  \overline{z}\right)  $ are given by $b_{n}=\frac{a_{n}%
}{n\left(  n^{2}-1\right)  }.$ From here we get that Sobolev $3\left/
2\right.  $ norm of $\Psi\left(  \overline{z}\right)  \left\vert _{S^{1}%
}\right.  $ is given by
\[
\left\Vert \Psi\right\Vert _{\mathbf{H}_{S^{1},h}^{3\left/  2\right.  }}^{2}=%
{\displaystyle\sum\limits_{n=2}^{\infty}}
\left(  n\left(  n^{2}-1\right)  \right)  \left\vert b_{n}\right\vert ^{2}=%
{\displaystyle\sum\limits_{n=2}^{\infty}}
\frac{n\left(  n^{2}-1\right)  \left\vert a_{n}\right\vert ^{2}}{\left(
n\left(  n^{2}-1\right)  \right)  ^{2}}=
\]%
\[%
{\displaystyle\sum\limits_{n=2}^{\infty}}
\frac{\left\vert a_{n}\right\vert ^{2}}{n\left(  n^{2}-1\right)  }=\left\Vert
\Psi(z)\right\Vert _{\mathbf{H}_{S^{1},h}^{3\left/  2\right.  }}%
^{2}=\left\Vert \psi(z)\right\Vert _{\mathbf{H}_{S^{1},h}^{-3\left/  2\right.
}}^{2}<\infty.
\]
Lemma \ref{Pair0} is proved. $\blacksquare$

\begin{lemma}
\label{SS}Let $\psi_{i}\left(  z\right)  $ be two holomorphic functions in
$\mathbb{D}^{\ast}$ such that
\[
\psi_{i}\left(  z\right)  \left\vert _{S^{1}}\right.  =\psi_{i}(\theta)=%
{\displaystyle\sum\limits_{n=2}^{\infty}}
a_{n}^{i}e^{-2\pi i(n-2)\theta}\in\mathbf{H}_{S_{1},h}^{-3\left/  2\right.
}.
\]
Let $\psi_{i}(\overline{z})=%
{\displaystyle\sum\limits_{n=2}^{\infty}}
a_{n}^{i}\left(  \overline{z}\right)  ^{n-2}$ be the anti holomorphic
functions on $\mathbb{D}$ such that $\psi_{i}\left(  \overline{z}\right)
\left\vert _{S^{1}}\right.  =\psi_{i}(\theta).$ Then the following formulas
hold:%
\begin{equation}
\left\langle \psi_{1}(\theta),\psi_{2}(\theta)\right\rangle _{\mathbf{H}%
_{S^{1},h}^{-3\left/  2\right.  }}:=\frac{1}{2\pi}%
{\displaystyle\int\limits_{\mathbb{D}}}
(1-|z|^{2})^{2}\psi_{1}(\overline{z})\overline{\psi_{2}(\overline{z})}%
dz\wedge\overline{dz}<\infty\label{pl2}%
\end{equation}

\end{lemma}

\textbf{Proof: }The definition of the Hardy space $\mathbf{H}_{S_{1}%
,h}^{-3\left/  2\right.  }$ implies
\begin{equation}
\psi_{i}\left(  z\right)  \in\mathbf{H}_{S_{1},h}^{-3\left/  2\right.
}\Longleftrightarrow\left\Vert \psi_{i}\right\Vert _{\mathbf{H}_{S^{1}%
,h}^{-3\left/  2\right.  }}^{2}=%
{\displaystyle\sum\limits_{n=2}^{\infty}}
\frac{\left\vert a_{n}^{i}\right\vert ^{2}}{n\left(  n^{2}-1\right)  }%
<\infty\label{h3/2}%
\end{equation}
for $i=1$ and $2$. By the definition of the scalar product in the Hilbert
space $\mathbf{H}_{S^{1},h}^{-3\left/  2\right.  }$ we have
\begin{equation}
\left\langle \psi_{1}(\theta),\overline{\psi_{2}(\theta)}\right\rangle
_{\mathbf{H}_{S^{1},h}^{-3\left/  2\right.  }}=%
{\displaystyle\sum\limits_{n=2}^{\infty}}
\frac{a_{n}^{1}\overline{a_{n}^{2}}}{n\left(  n^{2}-1\right)  }. \label{pl20}%
\end{equation}
Direct computations show that%
\[
\frac{1}{2\pi}%
{\displaystyle\int\limits_{\mathbb{D}}}
(1-|z|^{2})^{2}\psi_{1}(\overline{z})\overline{\psi_{2}(\overline{z})}%
dz\wedge\overline{dz}=
\]%
\begin{equation}%
{\displaystyle\int\limits_{0}^{1}}
\left(  1-r^{2}\right)  ^{2}\left(
{\displaystyle\sum\limits_{n=2}^{\infty}}
a_{n}^{1}\overline{a_{n}^{2}}r^{2n-1}\right)  dr=%
{\displaystyle\sum\limits_{n=2}^{\infty}}
\frac{a_{n}^{1}\overline{a_{n}^{2}}}{n\left(  n^{2}-1\right)  }. \label{pl2a}%
\end{equation}
The facts that $\psi_{1}\left(  z\right)  $ \& $\psi_{2}\left(  z\right)
\left\vert _{S^{1}}\right.  \in\mathbf{H}_{S_{1},h}^{-3\left/  2\right.  }$
and Cauchy-Schwarz inequality imply that the power series in $\left(
\ref{pl2a}\right)  $ converges. Comparing formula $\left(  \ref{pl2a}\right)
$ with $\left(  \ref{pl20}\right)  $ we derive formula $\left(  \ref{pl2}%
\right)  $. Lemma \ref{SS} is proved. $\blacksquare$

\begin{lemma}
\label{SSa}\textit{Let }$\psi(z)$ be\ a holomorphic function in $\mathbb{D}$
such that $\psi\left(  z\right)  \in\mathbf{H}_{S_{1},h}^{-3\left/  2\right.
}.$ \textit{Then for each holomorphic function }$\eta\left(  z\right)
$\textit{ on} $\mathbb{D}^{\ast}$ \textit{such that} $\eta\left(  z\right)
\left\vert _{S^{1}}\right.  \in\mathbf{H}_{S_{1}}^{-3\left/  2\right.  }$
\textit{and}$\mathit{\ }\left\Vert \eta\right\Vert _{\mathbf{H}_{S^{1}%
,h}^{-3\left/  2\right.  }}^{2}=1,$ \textit{we have }%
\begin{equation}
\underset{\left\Vert \eta\right\Vert _{\mathbf{H}_{S^{1},h}^{-3\left/
2\right.  }}^{2}=1}{\max}\left\vert L_{\psi}\left(  \eta\right)  \right\vert
=\left\langle \Psi\left(  \theta\right)  ,\psi(\theta)\right\rangle
_{\mathbf{L}^{2}\left(  S^{1}\right)  }=\left\Vert \psi\right\Vert
_{\mathbf{H}_{S^{1},h}^{-3\left/  2\right.  }}^{2}=\left\Vert \Psi\right\Vert
_{\mathbf{H}_{S^{1},h}^{3\left/  2\right.  }}^{2}. \label{H10}%
\end{equation}

\end{lemma}

\textbf{Proof: } $\left(  \ref{h3/2}\right)  $, $\left(  \ref{pl2a}\right)
,$\ the definition $L_{\psi}(\eta(\theta))=\left\langle \psi,\eta\right\rangle
_{\mathbf{H}_{S^{1},h}^{-3\left/  2\right.  }}$ and Cauchy-Schwarz inequality
imply that
\[
\left\vert L_{\psi}(\eta(\theta))\right\vert =\left\vert \left\langle
\psi,\eta\right\rangle _{\mathbf{H}_{S^{1},h}^{-3\left/  2\right.  }%
}\right\vert ^{2}\leq\left\Vert \psi\right\Vert _{\mathbf{H}_{S^{1}%
,h}^{-3\left/  2\right.  }}^{2}\left\Vert \eta\right\Vert _{\mathbf{H}%
_{S^{1},h}^{-3\left/  2\right.  }}^{2}.
\]
Since we assumed that $\left\Vert \eta\right\Vert _{\mathbf{H}_{S^{1}%
,h}^{-3\left/  2\right.  }}^{2}=1$ then
\[
\left\vert L_{\psi}(\eta(\theta))\right\vert =\left\vert \left\langle
\psi,\eta\right\rangle _{\mathbf{H}_{S^{1},h}^{-3\left/  2\right.  }%
}\right\vert ^{2}\leq\left\Vert \psi\right\Vert _{\mathbf{H}_{S^{1}%
,h}^{-3\left/  2\right.  }}^{2}.
\]
Let $\eta=\frac{\psi}{\left\Vert \psi\right\Vert }.$ So we have
\[
\left\vert L_{\psi}\left(  \frac{\psi}{\left\Vert \psi\right\Vert }%
(\theta)\right)  \right\vert =\left\vert \left\langle \psi,\frac{\psi
}{\left\Vert \psi\right\Vert }\right\rangle _{\mathbf{H}_{S^{1},h}^{-3\left/
2\right.  }}\right\vert ^{2}=\left\Vert \psi\right\Vert _{\mathbf{H}_{S^{1}%
,h}^{-3\left/  2\right.  }}^{2}.
\]
The definition $\left(  \ref{2a}\right)  $ of $\Psi(z)$ implies that%
\[
\left\Vert \left.  \Psi(z)\right\vert _{S^{1}}\right\Vert _{\mathbf{H}%
_{S^{1},h}^{3\left/  2\right.  }}^{2}=%
{\displaystyle\sum\limits_{n\geq1}}
\frac{\left\vert a_{n}\right\vert ^{2}}{n(n+1)(n+2)}=\left\Vert \psi
\right\Vert _{\mathbf{H}_{S^{1},h}^{-3\left/  2\right.  }}^{2}.
\]
Lemma \ref{SSa} is proved. $\blacksquare$

According to Lemma \ref{SSa} we have%
\[
\underset{\left\Vert \psi_{2}\right\Vert _{\mathbf{H}_{S^{1},h}^{-3\left/
2\right.  }}^{2}=1}{\max}\left\vert L_{\psi_{1}}(\psi_{2}(\theta))\right\vert
=\left\Vert \psi_{1}\right\Vert _{\mathbf{H}_{S^{1},h}^{-3\left/  2\right.  }%
}^{2}.
\]
Formula $\left(  \ref{pl2}\right)  ,$ the expressions for $\Psi\left(
\theta\right)  $ given by $\left(  \ref{3a}\right)  $ and \ $\eta(\overline
{z})\left\vert _{S^{1}}\right.  \in\mathbf{H}_{S_{1},h}^{-3\left/  2\right.
}$ imply that%
\begin{equation}
\left\langle \Psi\left(  \theta\right)  ,\eta\left(  \theta\right)
\right\rangle _{\mathbf{L}^{2}(S^{1})}=\left\langle \psi,\eta\right\rangle
_{\mathbf{H}_{S^{1},h}^{-3\left/  2\right.  }}=%
{\displaystyle\sum\limits_{n>1}}
\frac{a_{n}\overline{b_{n}}}{n(n^{2}-1)}.
\end{equation}
This fact implies that the linear functional $L_{\psi}$ defined by $\left(
\ref{lf}\right)  $ is continuous on $\mathbf{H}_{S^{1},h}^{-3\left/  2\right.
}.$ Thus $L_{\psi}\in\left(  \mathbf{H}_{S^{1},h}^{-3\left/  2\right.
}\right)  ^{\ast}=\mathbf{H}_{S^{1},h}^{3\left/  2\right.  }$ and the map%
\begin{equation}
\psi\in\mathbf{H}_{S^{1},h}^{-3\left/  2\right.  }\rightarrow L_{\psi}%
\in\left(  \mathbf{H}_{S^{1},h}^{-3\left/  2\right.  }\right)  ^{\ast
}=\mathbf{H}_{S^{1},h}^{3\left/  2\right.  } \label{DM}%
\end{equation}
is linear and continuous. Formula $\left(  \ref{pl2a}\right)  $ implies that
since we can choose $\psi_{1}=\psi_{2},$ the map $\left(  \ref{DM}\right)  $
has zero kernel. Part \textbf{1} and \textbf{2} of Theorem \ref{Pair} are
proved. $\blacksquare$

Let $\eta(\overline{z})$ be any antiholomorphic function on $\mathbb{D}$ such
that
\[
\eta(\overline{z})\left\vert _{S^{1}}\right.  =\text{ }\eta\left(
\theta\right)  =%
{\displaystyle\sum\limits_{n>1}}
b_{n}e^{-i\left(  n-2\right)  \theta}\in\mathbf{H}_{S_{1},h}^{-3\left/
2\right.  }.
\]
Formula $\left(  \ref{pl2}\right)  ,$ the expressions for $\Psi\left(
\theta\right)  $ given by $\left(  \ref{3a}\right)  $ and \ $\eta(\overline
{z})\left\vert _{S^{1}}\right.  \in\mathbf{H}_{S_{1},h}^{-3\left/  2\right.
}$ imply that%
\begin{equation}
\left\langle \Psi\left(  \theta\right)  ,\eta\left(  \theta\right)
\right\rangle _{\mathbf{L}^{2}(S^{1})}=\left\langle \psi,\eta\right\rangle
_{\mathbf{H}_{S^{1},h}^{-3\left/  2\right.  }}=%
{\displaystyle\sum\limits_{n>1}}
\frac{a_{n}\overline{b_{n}}}{n(n^{2}-1)}. \label{l2c}%
\end{equation}
Comparing $\left(  \ref{pl2}\right)  $ and $\left(  \ref{l2c}\right)  $ we get
that
\begin{equation}
\left\langle \psi\left(  \theta\right)  ,\eta\left(  \theta\right)
\right\rangle _{\mathbf{H}_{S^{1},h}^{-3\left/  2\right.  }}=%
{\displaystyle\int\limits_{\mathbb{D}}}
(1-|z|^{2})^{2}\psi(\overline{z})\overline{\eta(\overline{z})}dz\wedge
\overline{dz}=\left\langle \Psi\left(  \theta\right)  ,\eta\left(
\theta\right)  \right\rangle _{\mathbf{L}^{2}(S^{1})}. \label{SP}%
\end{equation}
Thus part \textbf{3} of Theorem \ref{Pair} is proved. $\blacksquare$ Theorem
\ref{Pair} is proved. $\blacksquare$

\subsection{Geometric Interpretation of the Duality}

The motivation for the proof of Theorem \ref{Co} is the generalization of the
following fact about the finite dimensional Teichm\"{u}ller Theory to the
infinite dimensions.

Let $R$ be a Riemann surface of genus $g$ greater than, or equal to two. Let
$g_{0}$ be the metric with constant curvature in $R.$ Let $\mathbf{T}_{g}$ be
the Teichm\"{u}ller space of all Riemann surfaces of genus $g.$ It is a well
known fact that the holomorphic tangent space $T_{\tau_{R},\mathbf{T}_{g}}$ is
isomorphic to the space $\mathbb{H}^{1}\left(  R,T_{R}^{1,0}\right)  $ of
harmonic $(0,1)$ forms with coefficients in the holomorphic tangent bundle
with respect the metric of constant curvature on $R.$ The cotangent space
$\Omega_{\tau_{R},\mathbf{T}_{g}}^{1,0}$ is isomorphic to the space
$H^{0}\left(  R,\left(  \Omega_{R}^{1,0}\right)  ^{\otimes2}\right)  $ of
holomorphic quadratic differentials. See \cite{Ahlfors}.

The Serre duality $\left(  H^{0}\left(  R,\left(  \Omega_{R}^{1,0}\right)
^{\otimes2}\right)  \right)  ^{\ast}=\mathbb{H}^{1}\left(  R,T_{R}%
^{1,0}\right)  ,$ is given explicitly as follows; Let $z=x+iy,$
\[
\psi(z)\left(  dz\right)  ^{\otimes2}\in H^{0}\left(  R,\left(  \Omega
_{R}^{1,0}\right)  ^{\otimes2}\right)
\]
and $\frac{4dx\otimes dy}{\left(  1-x^{2}-y^{2}\right)  ^{2}}$ be the metric
with constant curvature on $\mathbb{D},$ which is the universal cover of $R$.
Then%
\[
\overline{\left(  \psi(z)\left(  dz\right)  ^{\otimes2}\right)  \lrcorner
\left(  \left(  1-\left\vert z\right\vert ^{2}\right)  ^{2}\frac{d}{dz}%
\otimes\overline{\frac{d}{dz}}\right)  }=
\]%
\[
\left(  1-\left\vert z\right\vert ^{2}\right)  ^{2}\overline{\psi(z)}\left(
\overline{dz}\otimes\frac{d}{dz}\right)  \in\mathbb{H}^{1}\left(
R,T_{R}^{1,0}\right)  .
\]
will be a harmonic element of $\mathbb{H}^{1}\left(  R,T_{R}^{1,0}\right)  $
with respect to $g_{0}.$

\begin{theorem}
\label{Co}The completion of $T_{id,\mathbf{T}^{\infty}}$ with respect to the
left invariant K\"{a}hler metric on $\mathbf{T}^{\infty}$ is a Hilbert space
$\left(  T_{id,\mathbf{T}^{\infty}}\right)  ^{3/2}$ isomorphic to to
$\mathbf{H}_{S^{1},h}^{3/2}.$ The completion of $\Omega_{id,\mathbf{T}%
^{\infty}}^{1,0}$ with respect to the left invariant K\"{a}hler metric on
$\mathbf{T}^{\infty}$ is a Hilbert space $\left(  \Omega_{id,\mathbf{T}%
^{\infty}}^{1,0}\right)  ^{-3/2}$ isomorphic to to $\mathbf{H}_{S^{1}%
,h}^{-3/2}.$
\end{theorem}

\textbf{Proof: }The proof of Theorem \ref{Co} follows from the following Lemma:

\begin{lemma}
\label{Co1}\textbf{1. }The holomorphic cotangent bundle $\Omega_{id,\mathbf{T}%
^{\infty}}^{1}$ at $id\in\mathbf{T}^{\infty}$ is canonically isomorphic to the
space%
\begin{equation}
\left\{  \left.  \psi\left(  \overline{z}\right)  \left(  dz\right)
^{\otimes2}\right\vert \psi\left(  \overline{z}\right)  =%
{\displaystyle\sum\limits_{n\geq2}}
a_{n}\left(  \overline{z}\right)  ^{n}\text{ and }|z|<1\right\}  , \label{AA0}%
\end{equation}
where $\left.  \psi\left(  \overline{z}\right)  \right\vert _{S^{1}}$ is a
$C^{\infty}$ function. \textbf{2. }The holomorphic tangent space
$T_{id,\mathbf{T}^{\infty}}$ at $id\in\mathbf{T}^{\infty}$ can be identified
with all Beltrami differentials%
\begin{equation}
(1-|z|^{2})^{2}\psi\left(  \overline{z}\right)  \left(  \overline{dz}%
\otimes\frac{d}{dz}\right)  =\mu_{\psi}\left(  \overline{dz}\otimes\frac
{d}{dz}\right)  , \label{AA1}%
\end{equation}
where $\left.  \psi\left(  \overline{z}\right)  \right\vert _{S^{1}}$ is a
$C^{\infty}.$
\end{lemma}

\textbf{Proof of part 1:} According to \cite{psegal} the space $\mathbf{T}%
^{\infty}$ can be viewed as the deformation space of all complex structures on
the unit disk $\mathbb{D}.$ The holomorphic cotangent bundle $\Omega
_{id,\mathbf{T}^{\infty}}^{1}$ at $id\in\mathbf{T}^{\infty}$ can be identified
with the space of all holomorphic quadratic differentials
\begin{equation}
\psi\left(  \frac{1}{z}\right)  \left(  d\left(  \frac{1}{z}\right)  \right)
^{\otimes2}=z^{4}\psi\left(  \frac{1}{z}\right)  \left(  dz\right)
^{\otimes2}, \label{AA}%
\end{equation}
where $\psi\left(  \frac{1}{z}\right)  $ is the Schwarzian of a univalent
function $\Phi\left(  \frac{1}{z}\right)  $ on the complement of the unit
disk. This follows from Theorem \ref{AW}. So we have
\[
\psi\left(  \frac{1}{z}\right)  =%
{\displaystyle\sum\limits_{n\geq4}}
a_{n}\left(  \frac{1}{z}\right)  ^{n}%
\]
and it is a complex analytic function in $\mathbb{C}-\mathbb{D}=\mathbb{D}%
^{\ast}$ such that $\left.  \psi\left(  \frac{1}{z}\right)  \right\vert
_{S^{1}}$ is a $C^{\infty}.$ We can canonically continue $\left.  \psi\left(
\frac{1}{z}\right)  \right\vert _{S^{1}}$ to an antiholomorphic function
\begin{equation}
\psi\left(  \overline{z}\right)  =%
{\displaystyle\sum\limits_{n>0}}
a_{n}\left(  \overline{z}\right)  ^{n} \label{AA2}%
\end{equation}
in $\mathbb{D}.$ Thus $\left(  \ref{AA}\right)  $ and $\left(  \ref{AA2}%
\right)  $ imply $\left(  \ref{AA0}\right)  .$ $\blacksquare$

\textbf{Proof of part 2: }According to \cite{witten2} the left invariant
K\"{a}hler metric on $\mathbf{T}^{\infty}$ is defined as follows on
$\Omega_{id,\mathbf{T}^{\infty}}^{1}:$%
\[
\frac{\sqrt{-1}}{2\pi}%
{\displaystyle\int\limits_{\mathbb{D}}}
\psi_{1}\left(  \overline{z}\right)  \left(  dz\right)  ^{\otimes2}%
\overline{\psi_{2}\left(  \overline{z}\right)  }\left(  \overline{dz}\right)
^{\otimes2}\left(  1-|z|^{2}\right)  ^{2}\frac{d}{dz}\wedge\overline{\frac
{d}{dz}}=
\]%
\[
\frac{\sqrt{-1}}{2\pi}%
{\displaystyle\int\limits_{\mathbb{D}}}
\left(  1-|z|^{2}\right)  ^{2}\psi_{1}\left(  \overline{z}\right)
\overline{\psi_{2}\left(  \overline{z}\right)  }dz\wedge\overline{dz}.
\]
Since we know that the metric gives a canonical isomorphism between
$\Omega_{id,\mathbf{T}^{\infty}}^{1}$ the conjugate of the tangent space
$T_{id,\mathbf{T}^{\infty}}$ we derive that$T_{id,\mathbf{T}^{\infty}}$ can be
identified with all Beltrami differentials%
\[
(1-|z|^{2})^{2}\psi\left(  \overline{z}\right)  \left(  \overline{dz}%
\otimes\frac{d}{dz}\right)  =\mu_{\psi}\left(  \overline{dz}\otimes\frac
{d}{dz}\right)  ,
\]
where $\left.  \psi\left(  \overline{z}\right)  \right\vert _{S^{1}}$ are
$C^{\infty}$ functions$.$ So $\left(  \ref{AA1}\right)  $ is proved.
$\blacksquare$

Theorem \ref{Pair} and Lemma \ref{Co1} imply Theorem \ref{Co}. $\blacksquare$

\begin{corollary}
\label{pair1}Let $|z|<1,$ $\psi(z)=%
{\displaystyle\sum\limits_{n=2}}
a_{n}\left(  \frac{1}{z}\right)  ^{n-2}$ be a holomorphic function on
$\mathbb{D}^{\ast}$. We already proved that the quadratic differential
$\psi(z)d\left(  \frac{1}{z}\right)  ^{\otimes2}\in\Omega_{id,\mathbf{T}%
^{\infty}}^{1,0}.$ Then the Weil-Petersson norm of the quadratic differential
$\psi(z)d\left(  \frac{1}{z}\right)  ^{\otimes2}$ with respect to the left
invariant K\"{a}hler metric is given by
\[
\left\Vert \psi\right\Vert _{W.P.}^{2}=L_{\psi}\left(  \psi\right)
=\left\langle \Psi\left(  \theta\right)  ,\psi\left(  \theta\right)
\right\rangle _{\mathbf{L}^{2}(S^{1})}=\left\Vert \psi\left(  \theta\right)
\right\Vert _{\mathbf{H}_{S^{1}}^{-3/2}}^{2}.
\]

\end{corollary}

\begin{corollary}
\label{CS3/2}Let $\mu_{\psi}\in$ $\mathbb{H}^{-1,1}(\mathbb{D}),$ where
$\mathbb{H}^{-1,1}(\mathbb{D})$ is defined by $\left(  \ref{2a}\right)  .$
Then the map $\digamma$ defined by $\left(  \ref{5a}\right)  $ is surjective
isomorphism between the Hilbert spaces $\mathbb{H}^{-1,1}(\mathbb{D})$ and
$\mathbf{H}_{h,S^{1}}^{3/2}$
\end{corollary}

\textbf{Proof: }The definition $\left(  \ref{2a}\right)  $ of $\mathbb{H}%
^{-1,1}(\mathbb{D}),$ the definition $\left(  \ref{3a}\right)  $ and the
definition $\left(  \ref{5a}\right)  $ of the map $\digamma$ directly imply
that the map $\digamma$ is surjective one to one. Part 1 of Corollary
\ref{CS3/2} is proved. $\blacksquare$

\begin{remark}
\label{4}In the paper \cite{TT} on page 23 it is stated "\textit{In this
section we are going to endow }$T(1)$ (the universal Teichm\"{u}ller space"
with a structure of a complex manifold modeled on the separable Hilbert space
\textit{ }%
\[
A_{2}(\mathbb{D}):=\left\{  \phi\text{ hol in }\mathbb{D}:\left\Vert
\phi\right\Vert _{2}^{2}=%
{\displaystyle\iint\limits_{\mathbb{D}}}
\left\vert \phi\right\vert ^{2}\left(  1-\left\vert z\right\vert ^{2}\right)
^{2}dz^{2}\right\}
\]
\textit{of holomorphic function on }$D.$ \textit{In the corresponding
topology, the universal Teichm\"{u}ller space T(1) is a disjoint union of
uncountably many components on which the right translations act transitively}%
". Later on page the authors wrote a paragraph; "\textbf{4.2. Weil-Petersson
metric on the universal Teichm\"{u}ller space.} In this section we consider
T(1) as a Hilbert Manifold. The Weil-Petersson metric on T(1) is a Hermitian
metric defined by the Hilbert space inner product on tangent spaces, which are
identified with the Hilbert space $H^{-1,1}\left(  \mathbb{D}^{\ast}\right)  $
by right translations (see Section 3.3). Thus the Weil-Petersson metric is a
right-invariant metric on T(1) defined at the origin of T(1) by%
\begin{equation}
\left\langle \mu,\nu\right\rangle _{W.P.}:=%
{\displaystyle\iint\limits_{\mathbb{D}}}
\mu\overline{\nu}\left(  1-\left\vert z\right\vert ^{2}\right)  ^{2}%
dz^{2},\text{ }\mu,\nu\in H^{-1,1}\left(  \mathbb{D}^{\ast}\right)
=T_{0}T(1). \tag{4.2}%
\end{equation}
To every $\mu\in H^{-}1,1\left(  \mathbb{D}^{\ast}\right)  $ there corresponds
a vector field $\frac{\partial}{\partial\varepsilon_{\mu}}$, over $V_{0}$
given by (2.5)-(2.7). We set for every $\kappa\in V_{0}$,%
\begin{equation}
g_{\mu,\overline{\nu}}(\kappa):=\left\langle \frac{\partial}{\partial
\varepsilon_{\mu}},\frac{\partial}{\partial\varepsilon_{\nu}}\right\rangle
_{W.P.}=%
{\displaystyle\iint\limits_{\mathbb{D}}}
P\left(  R\left(  \mu,\kappa\right)  \right)  \overline{P\left(  R\left(
\mu,\kappa\right)  \right)  }dz^{2}. \tag{4.3}%
\end{equation}
This formula explicitly defines the Weil-Petersson metric on the coordinate
chart $V_{0}$." \ In \cite{TT} is claimed that that the inner product on
$A_{2}(\mathbb{D})$ given by
\begin{equation}
\left\Vert \psi(\overline{z})\right\Vert ^{2}:=\frac{1}{2\pi\sqrt{-1}}%
{\displaystyle\iint\limits_{\mathbb{D}}}
\left(  1-|z|^{2}\right)  ^{2}\left\vert \psi(\overline{z})\right\vert
^{2}dz\wedge\overline{dz}=%
{\displaystyle\sum\limits_{n>1}}
\frac{|a_{n-2}|^{2}}{n(n^{2}-1)} \label{tt}%
\end{equation}
induces the Weil-Petersson metric. This is clearly the Hilbert space of the
boundary values of holomorphic functions with finite $-3\left/  2\right.  $
Sobolev norm. The Weil-Petersson metric on the $\mathbf{T}^{\infty}$ should be
induced by the norm of the Hilbert space of the boundary values of holomorphic
functions with finite $3\left/  2\right.  $ Sobolev norm. See \cite{nagver},
\cite{witten1} and \cite{witten2}.
\end{remark}

\section{$3\left/  2\right.  $ Hilbert Complex Analytic Structure on
$\mathbf{T}^{\infty}$}

\subsection{Special Properties of Two Operators}

\begin{definition}
\label{B2}Let $h(z)\in\mathbf{L}^{p}\left(  \mathbb{C}\right)  ,$ define
\begin{equation}
P\left(  h(\left(  \varsigma\right)  \right)  :=-\frac{1}{\pi}%
{\displaystyle\int\limits_{\mathbb{R}}}
{\displaystyle\int\limits_{\mathbb{R}}}
h(z)\left(  \frac{1}{z-\varsigma}-\frac{1}{z}\right)  dxdy, \label{bel2}%
\end{equation}
where $z=x+iy.$ See \cite{Ahlfors} chapter V, part A.
\end{definition}

\begin{definition}
\label{B3}Let $h(z)\in C_{0}^{2}\left(  \mathbb{C}\right)  ,$ define
\begin{equation}
T\left(  h\left(  \varsigma\right)  \right)  :=\underset{\varepsilon
\rightarrow0}{\lim}\left(  -\frac{1}{\pi}%
{\displaystyle\int\limits_{\left\vert z-\varsigma\right\vert ^{2}>\varepsilon
}}
h(z)\left(  \frac{1}{z-\varsigma}\right)  ^{2}dxdy\right)  , \label{bel3}%
\end{equation}
where $z=x+iy.$ See \cite{Ahlfors} chapter V, part A.
\end{definition}

The following Lemmas were proved in \cite{Ahlfors} Chapter V, part A:

\begin{lemma}
\label{ahl1}Let $h(z)\in\mathbf{L}^{p}\left(  \mathbb{C}\right)  ,$ $p>2.$ The
function $Ph$ is continuous and satisfies H\"{o}lder condition, i.e.%
\[
\left\vert Ph\left(  \varsigma_{1}\right)  -Ph\left(  \varsigma_{2}\right)
\right\vert \leq K_{p}\left\Vert h\right\Vert _{p}\left\vert \varsigma
_{1}-\varsigma_{2}\right\vert ^{1-\frac{2}{p}}.
\]

\end{lemma}

\begin{lemma}
\label{ahl2}For $h\in$ $\mathbf{L}^{p}\left(  \mathbb{C}\right)  ,$ $p>2,$ the
following relations hold%
\begin{equation}
\left(  Ph\right)  _{\overline{z}}=h\text{ \& }\left(  Ph\right)  _{z}=Th.
\label{bel9}%
\end{equation}

\end{lemma}

\begin{theorem}
\label{ahl3}For $h\in$ $\mathbf{L}^{p},$ $p>1$ on $\mathbb{C},$ we have
\begin{equation}
\left\Vert Th\right\Vert _{p}\leq C_{p}\left\Vert h\right\Vert _{p}
\label{bel10}%
\end{equation}
and
\begin{equation}
\underset{p\rightarrow2}{\lim}C_{p}=1. \label{bel11}%
\end{equation}

\end{theorem}

\subsection{Basis of $\ker\left(  \overline{\partial}_{\mu_{\psi}}\right)  $
for $\mu_{\psi}\in$ $\mathbb{H}_{1}^{-1,1}(\mathbb{D})$}

Theorem \ref{B4} was proved in \cite{STZ}. The ideas of the proof of Theorem
\ref{B4} are the same as in \cite{STZ}. Some of the technical steps are
simplified in the present proof.

\begin{theorem}
\label{B4}Let $\psi(\overline{z})$ be an antiholomorphic function in the unit
disk such that
\[
\psi(\overline{z})\left\vert _{S^{1}}\right.  \in\mathbf{H}_{S^{1}%
,h}^{-3\left/  2\right.  }\text{ }\&\text{ }\left\Vert \psi(\overline
{z})\left\vert _{S^{1}}\right.  \right\Vert _{S^{1},h}^{-3\left/  2\right.
}\leq c<1.
\]
Let $\mu_{\psi}(z)=\left(  1-\left\vert z\right\vert ^{2}\right)  ^{2}%
\psi(\overline{z})\left(  \overline{dz}\otimes\frac{d}{dz}\right)
\in\mathbb{H}_{1}^{-1,1}\left(  \mathbb{D}\right)  $ be the Beltrami
differential defined associated with $\mu_{\psi}(z)=\left(  1-\left\vert
z\right\vert ^{2}\right)  ^{2}\psi(\overline{z}).$ We assumed that
\[
\left\Vert \mu_{\psi}(z)\right\Vert _{\mathbf{L}_{\infty}}\leq k<1.
\]
Let choose $p>2$ such that so that $kC_{p}<1,$ where $\,C_{p}$ is defined by
Theorem \ref{ahl3}$,$ Let $T$ be defined by $\left(  \ref{bel3}\right)  $ and
$\nu^{(n)}(z)$ be:%
\begin{equation}
\nu^{(n)}(z):=\sum_{m=0}^{\infty}T_{m}^{n}\left(  \mu_{\psi}(z)\right)  ,
\label{bel4}%
\end{equation}
where $T_{0}^{n}\left(  \mu_{\psi}\right)  =z^{n-1},$ $T_{1}^{n}\left(
\mu_{\psi}\right)  =T\left(  \mu_{\psi}(z)z^{n-1}\right)  ,$ $T_{m}^{n}\left(
\mu_{\psi}\right)  =T\left(  \mu_{\psi}T_{m-1}^{n}\left(  \mu_{\psi}\right)
\right)  $ for $m>0.$ Define%
\begin{equation}
w^{(n)}(z)=z^{n}+nP\left(  \left(  \mu_{\psi}\left(  \nu^{(n)}(z)+z^{n-1}%
\right)  \right)  \right)  . \label{bel4a}%
\end{equation}
Then for any integer $n>0~$we have that $w^{(n)}(z)$ is the unique solution of
the following problem:%
\begin{equation}
\left\{
\begin{array}
[c]{c}%
\left(  \overline{\partial}-\mu_{\psi}\partial\right)  \left(  w^{(n)}%
(z)\right)  =0,\text{ \ \ \ \ \ \ \ \ \ \ \ \ \ \ \ \ \ \ \ \ \ \ \ \ \ \ }\\
\frac{\partial}{\partial z}\left(  w^{(n)}(z)\right)  -nz^{n-1}=n\nu
^{(n)}(z)\in\mathbf{L}^{p}\left(  \mathbb{C}\right)  ,\\%
{\displaystyle\int\limits_{0}^{2\pi}}
w^{(n)}\left(  e^{i\theta}\right)  d\theta=0\text{
\ \ \ \ \ \ \ \ \ \ \ \ \ \ \ \ \ \ \ \ \ \ \ \ \ \ \ \ \ \ \ }%
\end{array}
\right.  . \label{bel6}%
\end{equation}

\end{theorem}

\textbf{Proof: }The proof of Theorem \ref{B4} follows the argument used in
Chapter \textbf{V} in \cite{Ahlfors}.

\begin{lemma}
\label{hol1b}$\nu^{(n)}(z)$ are well defined functions on $\mathbb{C}$ and
$\nu^{(n)}(z)\in\mathbf{L}^{p}\left(  \mathbb{C}\right)  .$
\end{lemma}

\textbf{Proof: }The linear operator $h\rightarrow T\left(  \mu_{\psi
}(z)h\right)  $ on $\mathbf{L}^{p}\left(  \mathbb{C}\right)  $ has a norm
$\leq kC_{p}<1.$ Therefore the series%
\begin{equation}
\nu^{(n)}(z)=\sum_{m=0}^{\infty}T_{m}^{n}\left(  \mu_{\psi}(z)\right)
\label{be13}%
\end{equation}
is converging in $\mathbf{L}^{p}\left(  \mathbb{C}\right)  .$ From here we
derive that $\nu^{(n)}(z)\in\mathbf{L}^{p}\left(  \mathbb{C}\right)  .$ Lemma
\ref{hol1b} is proved. $\blacksquare$

\begin{lemma}
\label{hol1a}We have
\begin{equation}
\nu^{(n)}(z)\in\mathbf{L}^{2}\left(  \mathbb{C}\right)  \text{ \& }T\left(
\mu_{\psi}(z)\left(  \nu^{(n)}(z)\right)  \right)  \in\mathbf{L}^{2}\left(
\mathbb{C}\right)  . \label{Ahl0}%
\end{equation}

\end{lemma}

\textbf{Proof: }According to Theorem \ref{SS0}, the function $\mu_{\psi}(z)$
is bounded in the unit disk. We extended $\mu_{\psi}(z)$ outside the unit disk
to be zero. Thus $n\mu_{\psi}(z)z^{n-1}$ is a bounded function on $\mathbb{C}$
and zero outside the unit disk. Thus $\mu_{\psi}(z)\in\mathbf{L}_{0}%
^{2}\left(  \mathbb{C}\right)  $. Ahlfors proved that the operator
$T\,\ $defined by $\left(  \ref{bel3}\right)  $ defines an isometry of
$\mathbf{L}_{0}^{2}\left(  \mathbb{C}\right)  .$ See Lemma \textbf{2} in
Chapter \textbf{V, }part\textbf{ A }in \cite{Ahlfors}. The definition of
$\nu^{(n)}(z)$ and the above property of $T$ implies that $\nu^{(n)}%
(z)\in\mathbf{L}^{2}\left(  \mathbb{C}\right)  .$ We have
\[
\left\Vert \mu_{\psi}(z)\right\Vert _{\mathbf{L}^{2}\left(  \mathbb{D}\right)
}^{2}=\frac{\sqrt{-1}}{2\pi}%
{\displaystyle\int\limits_{\mathbb{D}}}
\left(  \left(  1-\left\vert z\right\vert ^{2}\right)  ^{2}\left\vert
\psi(z)\right\vert \right)  ^{2}dz\wedge\overline{dz}.
\]
Applying Schwarz inequality and $\left\Vert \mu_{\psi}(z)\right\Vert
_{\mathbf{L}^{\infty}\left(  \mathbb{D}\right)  }\leq k<1$ we get that
\begin{equation}
\left\Vert \mu_{\psi}(z)\right\Vert _{\mathbf{L}^{2}\left(  \mathbb{D}\right)
}^{2}\leq\left(  \frac{\sqrt{-1}}{2\pi}%
{\displaystyle\int\limits_{\mathbb{D}}}
\left(  1-\left\vert z\right\vert ^{2}\right)  ^{2}\left\vert \psi
(z)\right\vert dz\wedge\overline{dz}\right)  ^{2}<2. \label{bel14}%
\end{equation}
Combining Theorem \ref{ahl3} with $\left(  \ref{bel3}\right)  $ and $\left(
\ref{bel14}\right)  $ we get $\left(  \ref{Ahl0}\right)  .$ $\blacksquare$

\begin{lemma}
\label{hol1d}Let $w^{(n)}(z)$ be defined by $\left(  \ref{bel4a}\right)  .$
Then \textbf{1) }$w^{(n)}(z)$ satisfies in $\mathbb{C}$ the equation:
\begin{equation}
\overline{\partial}w^{(n)}(z)=n\mu_{\psi}(z)\left(  \nu^{(n)}(z)+z^{n-1}%
\right)  , \label{hold1}%
\end{equation}
\textbf{2)} $w^{(n)}-z^{n}\in\mathbf{L}^{p}\left(  \mathbb{C}\right)  $ where
$p\geq2$ was already fixed$,$ and \textbf{3)} $\overline{\partial}%
w^{(n)}(z)\in\mathbf{L}^{2}\left(  \mathbb{C}\right)  $ and $w^{(n)}%
(z)-z^{n}\in\mathbf{L}^{2}\left(  \mathbb{C}\right)  .$
\end{lemma}

\textbf{Proof of part 1: }The definitions of $w^{(n)}=z^{n}+nP\left(
\mu_{\psi}\left(  \nu^{(n)}(z)\right)  +z^{n-1}\right)  $ and of $\nu
^{(n)}(z)$ given by $\left(  \ref{bel4}\right)  $ imply
\[
w^{(n)}\left(  z\right)  -z^{n}=nP\left(  \mu_{\psi}\left(  \nu^{(n)}%
(z)+z^{n-1}\right)  \right)  .
\]
Lemma \ref{ahl2} and $\overline{\partial}\circ P=id$ imply that
\[
\overline{\partial}w^{(n)}(z)=\overline{\partial}\left(  z^{n}+nP\left(
\mu_{\psi}\left(  \nu^{(n)}(z)+z^{n-1}\right)  \right)  \right)  =n\mu_{\psi
}(z)\left(  \nu^{(n)}(z)+z^{n-1}\right)  .
\]
Part \textbf{1} of Lemma \ref{hol1d} is proved. $\blacksquare$

\textbf{Proof of part 2: }Ahlfors proved in Part \textbf{V, B }in
\cite{Ahlfors} that if
\[
\mu_{\psi}\left(  \nu^{(n)}(z)\right)  \in\mathbf{L}^{p}\left(  \mathbb{C}%
\right)  ,
\]
then%
\[
P\left(  \mu_{\psi}\left(  \nu^{(n)}(z)+z^{n-1}\right)  \right)  \in
\mathbf{L}^{p}\left(  \mathbb{C}\right)
\]
Part \textbf{2} of Lemma \ref{hol1d} is proved. $\blacksquare$

\textbf{Proof of part 3: }Part \textbf{3} follows directly from $\left(
\ref{hold1}\right)  $ and $\left(  \ref{Ahl0}\right)  .$ $\blacksquare$

\begin{lemma}
\label{hol1}We have
\begin{equation}
\partial w^{(n)}(z)-nz^{n-1}=n\nu^{(n)}(z)\in\mathbf{L}^{2}\left(
\mathbb{C}\right)  . \label{holg}%
\end{equation}

\end{lemma}

\textbf{Proof: }We assumed that $w^{(n)}(z)$ is defined by $\left(
\ref{bel4a}\right)  .$ So
\[
\partial w^{(n)}(z)-nz^{n-1}=n\partial\left(  {P}\left(  \mu_{\psi}\left(
\nu^{(n)}(z)+z^{n-1}\right)  \right)  \right)  .
\]
Thus $\partial P=T$ implies
\[
\partial w^{(n)}-nz^{n-1}=n\partial\left(  {P}\left(  \mu_{\psi}\left(
\nu^{(n)}(z)+z^{n-1}\right)  \right)  \right)  =nT\left(  \mu_{\psi}\left(
\nu^{(n)}(z)+z^{n-1}\right)  \right)  .
\]
The definition of the function $\nu^{(n)}(z)$ implies that
\[
T\left(  \mu_{\psi}\left(  \nu^{(n)}(z)+z^{n-1}\right)  \right)  =\nu
^{(n)}(z).
\]
Thus
\begin{equation}
\partial w^{(n)}-nz^{n-1}=n\nu^{(n)}(z). \label{holH}%
\end{equation}
Since $\mu_{\psi}$ has a compact support and thus $\mu_{\psi}\left(  \nu
^{(n)}(z)\right)  ,$ then
\begin{equation}
\mu_{\psi}\left(  \nu^{(n)}(z)\right)  \in\mathbf{L}^{2}\left(  \mathbb{C}%
\right)  . \label{holA}%
\end{equation}
Lemma \ref{hol1} follows from $\left(  \ref{holH}\right)  ,$ $\left(
\ref{holA}\right)  ,$\ the definition $\left(  \ref{bel4}\right)  $ of
$\nu^{(n)}(z)$ and Theorem \ref{ahl3}. $\blacksquare$

\begin{lemma}
\label{hol} Let $w^{(n)}(z)$ be defined by $\left(  \ref{bel4a}\right)  .$ Let
$\Phi_{\infty}^{(n)}(z):=w^{(n)}(z)\left\vert _{\mathbb{D}^{\ast}}\right.  .$
Then $\Phi_{\infty}^{(n)}(z)$ is a holomorphic function given by
\begin{equation}
\Phi_{\infty}^{(n)}(z)=z^{n}+\sum_{i=1}^{\infty}b_{i}z^{-i}. \label{Hol}%
\end{equation}

\end{lemma}

\textbf{Proof: }Since $\mu_{\psi}=0$ in ${\mathbb{{C}}}-\mathbb{D}%
=\mathbb{D}^{\ast}$ Lemma \ref{hold} implies
\[
\overline{\partial}w^{(n)}(z)\left\vert _{\mathbb{D}^{\ast}}\right.
=n\overline{\partial}P\left[  \mu_{\psi}\left(  \nu^{(n)}+z^{n-1}\right)
\right]  \left\vert _{\mathbb{D}^{\ast}}\right.  =
\]%
\[
\mu_{\psi}(z)\left(  \nu^{(n)}+z^{n-1}\right)  \left\vert _{\mathbb{D}^{\ast}%
}\right.  =0.
\]
Thus
\[
\Phi_{\infty}^{(n)}(z):=w^{(n)}(z)\left\vert _{\mathbb{D}^{\ast}}\right.
=z^{n}+G(z)+%
{\displaystyle\sum\limits_{i=0}^{\infty}}
b_{i}z^{-i},
\]
where $G(z)$ is an analytic function defined in $\mathbb{C}$. Since by
\cite{Ahlfors}
\[
\partial_{z}w^{(n)}(z)-nz^{n-1}=n\nu^{(n)}(z)\in\mathbf{L}^{p}\left(
\mathbb{C}\right)
\]
we have $G(z)=A_{0}$ is a constant. So we we can choose $A_{0}=0.$ Thus
$\Phi_{\infty}^{(n)}(z)$ satisfies
\[
\Phi_{\infty}^{(n)}(z)=z^{n}+\sum_{i=0}^{\infty}b_{i}z^{-i}.
\]
Thus $%
{\displaystyle\int\limits_{S^{1}}}
w^{(n)}(z)\frac{dz}{z}=0.$ Cauchy's theorem implies$%
{\displaystyle\int\limits_{S^{1}}}
w^{(n)}(z)\frac{dz}{z}=A_{0}=0.$ So $w^{(n)}(z),$ satisfies all the three
conditions in (\ref{bel6}). Lemma \ref{hol} is proved. $\blacksquare$

\begin{corollary}
\label{hold}$w^{n}(z)$ satisfies the Beltrami equation, i.e.%
\[
\overline{\partial}w^{(n)}(z)-\mu_{\psi}(z)\partial w^{(n)}(z)=0.
\]

\end{corollary}

\textbf{Proof: }According to $\left(  \ref{hold1}\right)  $ we have
$\overline{\partial}w^{(n)}(z)=n\mu_{\psi}(z)\left(  \nu^{(n)}(z)\right)  .$
We defined
\[
w^{(n)}(z):=z^{n}+nP\left(  \mu_{\psi}(z)\left(  \nu^{(n)}(z)+z^{n-1}\right)
\right)
\]
Lemma \ref{hol1} implies%
\begin{equation}
\partial w^{(n)}(z)=nz^{n-1}+n\nu^{(n)}(z). \label{hold2}%
\end{equation}
Substituting $\left(  \ref{hold2}\right)  $ in $\left(  \ref{hold1}\right)  $
we get $\overline{\partial}w^{(n)}(z)-\mu_{\psi}(z)\partial w^{(n)}(z)=0.$
Corollary \ref{hold} is proved. $\blacksquare$

Next we need to prove uniqueness of the solutions of the Beltrami equation. If
the Beltrami equation has two solutions $w_{i}^{(n)}(z)$ that satisfy the
three conditions in (\ref{bel6}) then the difference $w_{2}^{(n)}%
(z)-w_{1}^{(n)}(z)=W^{(n)}(z)$ will be also a solution of the Beltrami
equation. Therefore $W^{(n)}\left\vert _{\mathbb{D}^{\ast}}\right.  $ will be
a holomorphic function in $\mathbf{L}^{p}.$ From here we deduce that
$W^{(n)}(\infty)=0$ and thus we found a bounded solution $W^{(n)}(z)$ of the
Beltrami equation in $\mathbb{P}^{1}\left(  \mathbb{C}\right)  .$ But the
Beltrami equation is an elliptic. So by maximum principle of elliptic
equations it follows that $W^{(n)}(z)$ is a constant and since $W^{(n)}%
(\infty)=0$ we get that it is zero. Theorem \ref{B4} is proved. $\blacksquare$

\begin{corollary}
\label{B41}The functions $w^{(n)}(z)$ are continuous functions on
$\mathbb{C}.$
\end{corollary}

\textbf{Proof:} Corollary \ref{B41} follows from Theorem \ref{B4} directly.
$\blacksquare$

\subsection{The Embedding of $\mathbb{H}_{1}^{-1,1}(\mathbb{D})$ into the
space of Hilbert-Schmidt Operators on $\mathbb{HS}(\mathbf{H}^{+}%
,\mathbf{H}^{-}).$}

We recall that for $\mathbf{H}_{1}$ and $\mathbf{H}_{2}$ Hilbert spaces, an
operator $T:\mathbf{H}_{1}\rightarrow\mathbf{H}_{2}$ is called Hilbert-Schmidt
if for any orthogonal basis $\{e_{i}\}_{i\in I}$ of $\mathbf{H}_{1}$ we have
\[
\sum_{i\in I}\left\Vert Te_{i}\right\Vert ^{2}<\infty.
\]
If this is the case for one basis of $\mathbf{H}_{1}$ it is also the case for
every basis of $\mathbf{H}_{1}$.

\begin{theorem}
\label{Pr}Let $\mu_{\psi}\in\mathbb{H}^{-1,1}(\mathbb{D})$ be a Beltrami
differential where $\mathbb{H}^{-1,1}(\mathbb{D})$ is defined by $\left(
\ref{2a}\right)  $. Let
\[
\left\Vert \mu_{\psi}\right\Vert _{\mathbf{L}^{\infty}\left(  \mathbb{D}%
\right)  }\leq k<1.
\]
Then
\[
\mathbf{W}_{\mu_{\psi}}:=\ker\overline{\partial}_{\mu_{\psi}}:=\left\{
f(z)\left\vert _{S^{1}}\right.  \in\mathbf{L}^{2}\left(  S^{1}\right)
\left\vert \overline{\partial}_{\mu_{\psi}}\left(  f(z)\right)  =0\text{
}\right.  \right\}
\]
\textit{is a closed Hilbert subspace in} $\mathbf{L}^{2}\left(  S^{1}\right)
$ and \textit{the projection operator} $\mathbf{pr}_{\mu_{\psi}}%
^{+}:\mathbf{W}_{\mu_{\psi}}\rightarrow\mathbf{H}^{+}$ is an isomorphism. See
\cite{STZ}.
\end{theorem}

\textbf{Proof: }First we will prove that

\begin{lemma}
\label{hilb}$\mathbf{W}_{\mu_{\psi}}$ is a closed Hilbert space in
$\mathbf{L}^{2}\left(  S^{1}\right)  .$
\end{lemma}

\textbf{Proof: }In \cite{Ahlfors} it is proved that $\left\{  \left(
w^{(1)}(z)\right)  ^{n}\left\vert _{S^{1}}\right.  \right\}  $ form an
orthonormal basis on $\mathbf{W}_{\mu_{\psi}}$ and any function $f(z)$ on
$\mathbb{C}$ that satisfies the Beltrami equation and has a finite
$\mathbf{L}^{2}$ norm when restricted on $S^{1}$ can be expressed as follows:
\[
f(z)\left\vert _{S^{1}}\right.  :=%
{\displaystyle\sum\limits_{m\geq0}}
a_{n}\left(  \left.  w^{(1)}(z)\right\vert _{s^{1}}\right)  ^{n}\text{ \& }%
{\displaystyle\sum\limits_{m\geq0}}
\left\vert a_{n}\right\vert ^{2}<\infty.
\]
So $\mathbf{W}_{\mu_{\psi}}:=\left.  \ker\overline{\partial}_{\mu_{\psi}%
}\right\vert _{s^{1}}$ is a closed Hilbert space in $\mathbf{L}^{2}\left(
S^{1}\right)  $. $\blacksquare$

Let $w^{(n)}(z)$ be defined by $\left(  \ref{bel4}\right)  $ and $\left(
\ref{bel4a}\right)  .$ We know by Lemma \ref{hol1d} that $w^{(n)}(z)-z^{n}%
\in\mathbf{L}^{2}\left(  \mathbb{C}\right)  .$ Thus $\left.  w^{(n)}%
(z)\right\vert _{s^{1}}\in\mathbf{L}^{2}(S^{1}).$ Theorem \ref{B4} implies
that
\[
\{\left.  w^{(n)}(z)\right\vert _{s^{1}},\text{ }n=1,...\}
\]
is a basis of the Hilbert space $\ker\overline{\partial}_{\mu_{\psi}%
}=\mathbf{W}_{\mu_{\psi}}.$ From the definition of $\left.  w^{(n)}%
(z)\right\vert _{s^{1}}$ and Lemma \ref{hol} it follows that $\mathbf{pr}%
_{\mu_{\psi}}^{+}\left(  \left.  w^{(n)}(z)\right\vert _{s^{1}}\right)
=\left.  z^{n}\right\vert _{s^{1}}$ for $n=1,...$ is an isomorphism between
Hilbert spaces. This proves Theorem \ref{Pr}. $\blacksquare$

\begin{theorem}
\label{PrB}Let $\mu_{\psi}\in\mathbb{H}^{-1,1}(\mathbb{D})$ be a Beltrami
differential where $\mathbb{H}^{-1,1}(\mathbb{D})$ is defined by $\left(
\ref{2a}\right)  $ and
\[
\left\Vert \mu_{\psi}\right\Vert _{\mathbf{L}^{\infty}\left(  \mathbb{D}%
\right)  }\leq k<1.
\]
The projection operator $\mathbf{pr}_{\mu_{\psi}}^{-}:\mathbf{W}_{\mu_{\psi}%
}\rightarrow\mathbf{H}^{+}$ is a Hilbert-Schmidt operator.
\end{theorem}

Theorem \ref{PrB} was proved in \cite{STZ}. The proof presented in this paper
is technically much simpler.

\textbf{Proof: }The proof that $\mathbf{pr}_{\mu_{\psi}}^{-}:\mathbf{W}%
_{\mu_{\psi}}\rightarrow\mathbf{H}^{-}$ is a Hilbert-Schmidt operator is based
on the following observation. According to Theorem \ref{Pr} $\mathbf{pr}%
_{\mu_{\psi}}^{+}:\mathbf{W}_{\mu_{\psi}}\rightarrow\mathbf{H}^{+}$ is an
isomorphism. Therefore if we prove that
\begin{equation}
\mathbf{pr}_{\mu_{\psi}}^{-}\left(  \mathbf{pr}_{\mu_{\psi}}^{+}\right)
^{-1}:\mathbf{H}^{+}\rightarrow\mathbf{H}^{-} \label{compr}%
\end{equation}
is Hilbert-Schmidt operator then it will follow that $\mathbf{pr}_{\mu_{\psi}%
}^{-}:\mathbf{W}_{\mu_{\psi}}\rightarrow\mathbf{H}^{+}$ is Hilbert-Schmidt
too. The proof of this fact is based on several Lemmas.

\begin{lemma}
\label{B4a}We have%
\[
\mathbf{pr}_{\mu_{\psi}}^{-}\left(  \left.  w^{(n)}(z)\right\vert _{s^{1}%
}\right)  =
\]%
\begin{equation}
n\left.  P\left[  \mu_{\psi}\left(  \left(  \sum_{m=1}^{\infty}T_{m}%
^{n}\left(  \mu_{\psi}\right)  +z^{n-1}\right)  \right)  \right]  \right\vert
_{S^{1}}=%
{\displaystyle\sum\limits_{n>0}}
a_{n}e^{-in\theta}. \label{bel7a}%
\end{equation}

\end{lemma}

\textbf{Proof: }According to Lemma \ref{hol} and Theorem \ref{B4} the
restriction of $w^{(n)}(z)$ on $\mathbb{D}^{\ast}$ is a complex analytic
function such that it is in $\mathbf{L}^{p}\left(  \mathbb{C}\right)  .$ So
$\left(  \ref{Hol}\right)  $ implies that
\begin{equation}
\left.  w^{(n)}(z)\right\vert _{\mathbb{D}^{\ast}}=z^{n}+\sum_{j>1}a_{j}%
z^{-j}. \label{bel8}%
\end{equation}
The space $H_{-}$ defined by $\left(  \ref{h-}\right)  $ is spanned by
$z^{-j}$ for $j>0.$ On the other hand Theorem \ref{B4} implies that $\left.
w^{(n)}(z)\right\vert _{s^{1}}$ are restriction of holomorphic $\mathbf{L}%
^{2}$ functions from $\mathbb{D}^{\ast}$ on $S^{1}$ and they form a basis of
$\mathbf{W}_{\mu_{\psi}}.$ Thus $\left(  \ref{bel8}\right)  $ implies that
\[
\mathbf{pr}_{\mu_{\psi}}^{-}\left(  \left.  w^{(n)}(z)\right\vert _{s^{1}%
}\right)  =\left.  w^{(n)}(z)\right\vert _{s^{1}}-\left.  z^{n}\right\vert
_{s^{1}}=%
{\displaystyle\sum\limits_{n>0}}
a_{n}e^{-in\theta}.
\]
Lemma \ref{B4a} implies that%
\[
\mathbf{pr}_{\mu_{\psi}}^{-}\left(  \left.  w^{(n)}(z)\right\vert _{s^{1}%
}\right)  =n\left.  P\left[  \mu_{\psi}\left(  \sum_{m=1}^{\infty}(T_{m}%
^{n}\left(  \mu_{\psi}\right)  +z^{n-1}\right)  \right]  \right\vert _{S^{1}%
}=
\]%
\begin{equation}%
{\displaystyle\sum\limits_{n>0}}
a_{n}e^{-in\theta}\in\mathbf{L}^{2}(S^{1}). \label{ant2}%
\end{equation}
Thus $\left(  \ref{bel7a}\right)  $ is proved. Lemma \ref{B4a} follows
directly. $\blacksquare$

\begin{lemma}
\label{B4b}Let $\left\langle \mathbf{pr}_{\mu_{\psi}}^{-}\left(
\mathbf{pr}_{\mu_{\psi}}^{+}\right)  ^{-1}\left(  \left.  z^{n}\right\vert
_{s^{1}}\right)  ,\left.  z^{-k}\right\vert _{s^{1}}\right\rangle $ denote the
scalar product in the Hilbert space $\mathbf{L}^{2}(S^{1}).$ Then we have%
\[%
{\displaystyle\sum\limits_{n=1}^{\infty}}
\left\Vert \mathbf{pr}_{\mu_{\psi}}^{-}\left(  \mathbf{pr}_{\mu_{\psi}}%
^{+}\right)  ^{-1}\left(  \left.  z^{n}\right\vert _{s^{1}}\right)
\right\Vert _{\mathbf{L}^{2}(S^{1})}^{2}=
\]%
\begin{equation}%
{\displaystyle\sum\limits_{n,k=1}^{\infty}}
\left\vert \left\langle \mathbf{pr}_{\mu_{\psi}}^{-}\left(  \mathbf{pr}%
_{\mu_{\psi}}^{+}\right)  ^{-1}\left(  \left.  z^{n}\right\vert _{s^{1}%
}\right)  ,\left.  z^{-k}\right\vert _{s^{1}}\right\rangle \right\vert
^{2}<\infty. \label{ant3}%
\end{equation}

\end{lemma}

\textbf{Proof: }We are going to use the convention that $z=r\exp(i\theta)$. An
elementary application of Stokes' theorem and $\left(  \ref{bel7a}\right)  $
gives that%
\[
\left\langle \mathbf{pr}_{\mu_{\psi}}^{-}\left(  \mathbf{pr}_{\mu_{\psi}}%
^{+}\right)  ^{-1}\left(  \left.  z^{n}\right\vert _{s^{1}}\right)  ,\left.
z^{-k}\right\vert _{s^{1}}\right\rangle =
\]%
\[
\frac{1}{2\pi}\int_{0}^{2\pi}\mathbf{pr}_{\mu_{\psi}}^{-}\left(
\mathbf{pr}_{\mu_{\psi}}^{+}\right)  ^{-1}\left(  \left.  z^{n}\right\vert
_{s^{1}}\right)  \overline{\exp(-ik\theta)}\ d\theta=
\]%
\[
\frac{1}{2\pi i}\int_{S^{1}}\mathbf{pr}_{\mu_{\psi}}^{-}\left(  \mathbf{pr}%
_{\mu_{\psi}}^{+}\right)  ^{-1}\left(  \left.  z^{n}\right\vert _{s^{1}%
}\right)  \left.  z^{k}\frac{dz}{z}\right\vert _{S^{1}}=
\]%
\[
\frac{1}{2\pi i}%
{\displaystyle\int\limits_{\mathbb{D}}}
\overline{\partial}\left(  \mathbf{pr}_{\mu_{\psi}}^{-}\left(  \mathbf{pr}%
_{\mu_{\psi}}^{+}\right)  ^{-1}\left(  z^{n}\right)  \right)  \wedge
(z^{k-1}dz)=
\]%
\begin{equation}
\frac{1}{2\pi i}%
{\displaystyle\int\limits_{\mathbb{D}}}
\overline{\partial}P\left[  n\mu_{\psi}\left(  \sum_{m=1}^{\infty}(T_{m}%
^{n}\left(  \mu_{\psi}\right)  +z^{n-1}\right)  \right]  \ \overline{dz}%
\wedge(z^{k-1}dz). \label{eq1}%
\end{equation}
But, from the construction of
\[
w^{(n)}=z^{n}+nP\left[  \mu_{\psi}\left(  \sum_{m=1}^{\infty}(T_{m}^{n}\left(
\mu_{\psi}\right)  +z^{n-1}\right)  \right]
\]
$\overline{\partial P}=id,$ and the definition of $\nu^{(n)}(z)$ we have that
\begin{equation}
\overline{\partial}\left(  P\left[  \mu_{\psi}\left(  \sum_{m=1}^{\infty
}(T_{m}^{n}\left(  \mu_{\psi}\right)  +z^{n-1}\right)  \right]  \right)
=\nu^{(n)}(z). \label{eq0}%
\end{equation}
So substituting $\left(  \ref{eq0}\right)  $ into $\left(  \ref{eq1}\right)  $
we get that%
\[
\left\langle \mathbf{pr}_{\mu_{\psi}}^{-}\left(  \mathbf{pr}_{\mu_{\psi}}%
^{+}\right)  ^{-1}\left(  \left.  z^{n}\right\vert _{s^{1}}\right)  ,\left.
z^{-k}\right\vert _{s^{1}}\right\rangle =
\]%
\begin{equation}
\frac{1}{2\pi i}%
{\displaystyle\int\limits_{\mathbb{D}}}
\left(  n\mu_{\psi}\left(  \sum_{m=1}^{\infty}T_{m}^{n}\left(  \mu_{\psi
}\right)  +z^{n-1}\right)  \right)  z^{k-1}dz\wedge\overline{dz}.
\label{eq:prep}%
\end{equation}
Cauchy-Schwarz (for the inner-product of $\mathbf{L}^{2}(\mathbb{D})$),
applied to the right hand side of equation (\ref{eq:prep}) gives:%
\[
\left\vert \left\langle \mathbf{pr}_{\mu_{\psi}}^{-}\left(  \mathbf{pr}%
_{\mu_{\psi}}^{+}\right)  ^{-1}\left(  \left.  z^{n}\right\vert _{s^{1}%
}\right)  ,\left.  z^{-k}\right\vert _{s^{1}}\right\rangle \right\vert ^{2}=
\]%
\[
\left\vert \frac{1}{2\pi i}%
{\displaystyle\int\limits_{\mathbb{D}}}
n\mu_{\psi}\left(  \sum_{m=1}^{\infty}T_{m}^{n}\left(  \mu_{\psi}\right)
+z^{n-1}\right)  z^{k-1}dz\wedge\overline{dz}\right\vert ^{2}\leq
\]%
\[
\left(  \frac{C}{2\pi i}%
{\displaystyle\int\limits_{\mathbb{D}}}
n^{2}\left\vert \sum_{m=1}^{\infty}T_{m}^{n}\left(  \mu_{\psi}\right)
+\mu_{\psi}z^{n-1}\right\vert ^{2}dz\wedge\overline{dz}\right)  \left(
\frac{1}{2\pi i}%
{\displaystyle\int\limits_{\mathbb{D}}}
\left(  \mu_{\psi}\right)  ^{2}\left\vert z\right\vert ^{2k-2}dz\wedge
\overline{dz}\right)  \leq
\]%
\[
n^{2}\left(  \left\Vert \sum_{m=1}^{\infty}T_{m}^{n}\left(  \mu_{\psi}\right)
\right\Vert _{\mathbf{L}^{2}(\mathbb{D})}^{2}+\frac{1}{2\pi i}%
{\displaystyle\int\limits_{\mathbb{D}}}
\left(  \mu_{\psi}\right)  ^{2}\left\vert z\right\vert ^{2n-2}\ d\bar{z}\wedge
dz\right)  \times
\]%
\begin{equation}
\frac{1}{2\pi i}%
{\displaystyle\int\limits_{\mathbb{D}}}
\left(  \mu_{\psi}\right)  ^{2}\left\vert z\right\vert ^{2k-2}dz\wedge
\overline{dz}. \label{D0}%
\end{equation}
We are going to estimate each of the norms of equation (\ref{D0}).

\begin{proposition}
\label{Eq1}The following inequality holds for $n>2$:%
\begin{equation}
\frac{1}{2\pi i}%
{\displaystyle\int\limits_{\mathbb{D}}}
\left(  \mu_{\psi}\right)  ^{2}\left\vert z\right\vert ^{2n-2}d\bar{z}\wedge
dz\leq\frac{C}{n\left(  n+1\right)  \left(  n+2\right)  \left(  n+3\right)  }.
\label{EQI}%
\end{equation}

\end{proposition}

\textbf{Proof of }$\left(  \ref{EQI}\right)  $\textbf{:} We have
\[
\frac{1}{2\pi i}%
{\displaystyle\int\limits_{\mathbb{D}}}
\left(  \mu_{\psi}\right)  ^{2}\left\vert z\right\vert ^{2n-2}d\bar{z}\wedge
dz=
\]%
\[
\frac{1}{2\pi}%
{\displaystyle\int\limits_{0}^{2\pi}}
{\displaystyle\int\limits_{0}^{1}}
(1-r^{2})^{4}\left\vert \psi\left(  \overline{z}\right)  \right\vert
^{2}r^{2n-1}\ dr\ d\theta=%
{\displaystyle\int\limits_{0}^{1}}
(1-r^{2})^{4}\left\vert \psi\left(  \overline{z}\right)  \right\vert
^{2}r^{2n-1}\ dr.
\]
By Theorem $\left(  \ref{SS1}\right)  $ we have $\left\vert \psi\left(
\left\vert z\right\vert \right)  \right\vert <C\left(  1-\left\vert
z\right\vert \right)  ^{-\alpha}$ for $0<\alpha<1.$ So
\[
\frac{1}{2\pi i}%
{\displaystyle\int\limits_{\mathbb{D}}}
\left(  \mu_{\psi}\right)  ^{2}\left\vert z\right\vert ^{2n-2}d\bar{z}\wedge
dz\leq%
{\displaystyle\int\limits_{0}^{1}}
\left(  (1-r^{2})^{2}\left\vert \psi\left(  \overline{z}\right)  \right\vert
\right)  ^{2}(1-r^{2})^{2}r^{2n-1}dr\leq
\]%
\[
c%
{\displaystyle\int\limits_{0}^{1}}
\left(  (1-r^{2})^{2(1-\alpha)}\right)  (1-r^{2})^{2}r^{2n-1}dr.
\]
Direct computations by integrations by parts and using that $0<\alpha<1$ and
thus $-1\leq1-2\alpha$ we get
\[%
{\displaystyle\int\limits_{0}^{1}}
(1-r^{2})^{4-2\alpha}r^{2n-1}dr=\frac{1}{2n}%
{\displaystyle\int\limits_{0}^{1}}
(1-r^{2})^{4-2\alpha}dr^{2n}=
\]%
\[
-\frac{1}{2n}%
{\displaystyle\int\limits_{0}^{1}}
r^{2n}d\left(  (1-r^{2})^{4-2\alpha}\right)  =-\frac{\left(  4-2\alpha\right)
}{2n}%
{\displaystyle\int\limits_{0}^{1}}
r^{2n+1}(1-r^{2})^{3-2\alpha}dr
\]%
\[
\frac{\left(  4-2\alpha\right)  }{2n}%
{\displaystyle\int\limits_{0}^{1}}
(1-r^{2})^{3-2\alpha}dr^{2n+2}=-\frac{\left(  4-2\alpha\right)  }{2n\left(
2n+2\right)  }%
{\displaystyle\int\limits_{0}^{1}}
(1-r^{2})^{3-2\alpha}dr^{2n+2}=
\]%
\[
\frac{\left(  4-2\alpha\right)  }{2n\left(  2n+2\right)  }%
{\displaystyle\int\limits_{0}^{1}}
r^{2n+2}d(1-r^{2})^{3-2\alpha}=\frac{\left(  4-2\alpha\right)  }{2n\left(
2n+2\right)  }%
{\displaystyle\int\limits_{0}^{1}}
r^{2n+3}(1-r^{2})^{2-2\alpha}dr.
\]
Integrating twice by parts the expression%
\[
\frac{\left(  4-2\alpha\right)  }{2n\left(  2n+2\right)  }%
{\displaystyle\int\limits_{0}^{1}}
r^{2n+3}(1-r^{2})^{2-2\alpha}dr
\]
we get%
\[
\frac{1}{2\pi i}%
{\displaystyle\int\limits_{\mathbb{D}}}
\left\vert \mu_{\psi}z\right\vert ^{2n-2}d\bar{z}\wedge dz=
\]%
\begin{equation}%
{\displaystyle\int\limits_{0}^{1}}
(1-r^{2})^{4-2\alpha}r^{2n-1}dr\leq\frac{C_{1}}{n\left(  n+1\right)  \left(
n+2\right)  \left(  n+3\right)  }. \label{eq4a}%
\end{equation}
$\left(  \ref{EQI}\right)  $ is proved. $\blacksquare$

\begin{proposition}
\label{Eq2}Let $\psi(\overline{z})=%
{\displaystyle\sum\limits_{p=2}}
a_{p}\overline{z}^{p-1}.$ Then%
\[
\left\Vert \sum_{m=0}^{\infty}T_{m}^{n}\left(  \mu_{\psi}\right)  \right\Vert
_{\mathbf{L}^{2}(\mathbb{D})}^{2}=
\]%
\[
\left\Vert \sum_{m=1}^{\infty}(T_{m}^{n}\left(  \mu_{\psi}\right)  +\mu_{\psi
}z^{n-1})\right\Vert _{\mathbf{L}^{2}(\mathbb{D})}^{2}\leq\left(  \frac{c_{0}%
}{1-c_{0}}+\left\Vert \mu_{\psi}z^{n-1}\right\Vert _{\mathbf{L}^{2}%
(\mathbb{D})}^{2}\right)  \leq
\]%
\begin{equation}
C_{0}%
{\displaystyle\sum_{p=2}^{\infty}}
\frac{1}{n\left(  n+1\right)  (n+2)(n+3)}<\infty. \label{EQII}%
\end{equation}

\end{proposition}

\textbf{Proof: }According to \cite{Ahlfors} the Hilbert transform $T$ is an
isometry of $\mathbf{L}^{2}({\mathbb{{C}}}).$ By the definition of the
operators $T_{m}^{n}(\mu_{\psi})=T\left(  \mu_{\psi}T_{m-1}^{n}(\mu_{\psi
})\right)  $and $T_{0}^{n}=\mu_{\psi}z^{n-1}$ and and the above results in
\cite{Ahlfors} imply that we have%
\[
\left\Vert T_{m}^{n}\right\Vert ^{2}\leq c_{0}\left\Vert T_{m-1}%
^{n}\right\Vert ^{2}\leq...\leq c_{0}^{n}\left\Vert \mu_{\psi}z^{n-1}%
\right\Vert _{\mathbf{L}^{2}(\mathbb{D})}^{2},
\]
where $c_{0}\overset{def}{=}\left\Vert \mu_{\psi}\right\Vert _{\mathbf{L}%
^{\infty}(\mathbb{D})}<1.$ Thus we have%
\begin{equation}
\left\Vert \sum_{m=1}^{\infty}(T_{m}^{n}\left(  \mu_{\psi}\right)  \right\Vert
_{\mathbf{L}^{2}(\mathbb{D})}^{2}\leq\frac{c_{0}}{1-c_{0}}. \label{d1}%
\end{equation}
Next we need to estimate
\[
\left\Vert \left(  \mu_{\psi}z^{n-1}\right)  \right\Vert _{\mathbf{L}%
^{2}(\mathbb{D})}^{2}=\frac{1}{2\pi\sqrt{-1}}%
{\displaystyle\int\limits_{\mathbb{D}}}
\left\vert \mu_{\psi}z^{n-1}\right\vert ^{2}dz\wedge\overline{dz}=
\]%
\[
\frac{1}{2\pi\sqrt{-1}}%
{\displaystyle\int\limits_{\mathbb{D}}}
\mu_{\psi}^{2}\left\vert z\right\vert ^{2n-2}dz\wedge\overline{dz}.
\]
We proved that $\mu_{\psi}$ is a bounded function on $\mathbb{C}.$ $\left(
\ref{EQI}\right)  $ implies%
\begin{equation}
\left\Vert \left(  \mu_{\psi}z^{n-1}\right)  \right\Vert _{\mathbf{L}%
^{2}(\mathbb{D})}^{2}\leq\frac{C}{n\left(  n+1\right)  \left(  n+2\right)
\left(  n+3\right)  }. \label{d0}%
\end{equation}%
\[
\left\Vert \sum_{m=0}^{\infty}(T_{m}^{n}\left(  \mu_{\psi}\right)  \right\Vert
_{\mathbf{L}^{2}(\mathbb{D})}^{2}\leq\frac{C_{1}}{n\left(  n+1\right)  \left(
n+2\right)  \left(  n+3\right)  }%
\]
Proposition \ref{Eq2} is proved. $\blacksquare$

To conclude the proof of Lemma \ref{B4b} we will use the two estimates
$\left(  \ref{EQI}\right)  $ and $\left(  \ref{EQII}\right)  $ above for the
norms that appear on the right hand side of equations $\left(  \ref{eq:prep}%
\right)  $ and $(\ref{D0})$ to get%
\[%
{\displaystyle\sum\limits_{n=1}^{\infty}}
\left\Vert \mathbf{pr}_{\mu_{\psi}}^{-}\left(  \mathbf{pr}_{\mu_{\psi}}%
^{+}\right)  ^{-1}\left(  z^{n}\right)  \right\Vert _{\mathbf{L}^{2}(S^{1}%
)}^{2}=%
{\displaystyle\sum\limits_{n,k=1}^{\infty}}
\left\vert \left\langle \left.  \nu^{(n)}(z)\right\vert _{s^{1}},\left.
z^{-k}\right\vert _{s^{1}}\right\rangle \right\vert ^{2}<
\]%
\[%
{\displaystyle\sum\limits_{k=1,n=1,m=1}^{\infty}}
\left(  \left\Vert \mu_{\psi}z^{k-1}\right\Vert _{\mathbf{L}^{2}(\mathbb{D}%
)}^{2}\right)  \times\left(  n^{2}\left\Vert (T_{m}^{n}\left(  \mu_{\psi
}\right)  )\right\Vert _{\mathbf{L}^{2}(\mathbb{D})}^{2}\right)  =
\]%
\[
C\left(  \sum_{k=1}^{\infty}\frac{1}{k\left(  k+1\right)  (k+2)(k+3)}\right)
\left(
{\displaystyle\sum_{n=1}^{\infty}}
\frac{n^{2}}{n\left(  n+1\right)  \left(  n+2\right)  \left(  n+3\right)
}\right)  <\infty.
\]
Lemma \ref{B4b} \ is proved. $\blacksquare$ Lemma \ref{B4b} implies Theorem
\ref{PrB}. $\blacksquare$

\subsection{Differential Geometry of the Segel Wilson Grassmannian}

\begin{definition}
\textbf{1. }The Segal-Wilson Grassmannian $\mathbb{G}r_{\infty}$\textbf{ }is
defined as the set of all closed subspaces $\mathbf{W}$ of $\mathbf{L}%
^{2}(S^{1})=\mathbf{H}^{+}\oplus\mathbf{H}^{-}$ such that the projection
\[
\mathrm{pr}_{+}:\mathbf{W}\rightarrow\mathbf{H}^{+}%
\]
is Fredholm and the projection
\[
\mathrm{pr}_{-}:\mathbf{W}\rightarrow\mathbf{H}^{-}%
\]
is Hilbert-Schmidt, where $\mathbf{H}$ is the $\mathbf{L}^{2}$ space of
complex functions on $S^{1}$. \textbf{2. }A linear operator $A\in
\mathbb{GL}_{\mathrm{res}}(\mathbf{L}^{2}(S^{1}))$ iff the following
conditions are satisfied: \textbf{A. }$A$ is an invertible bounded linear
operator of $\mathbf{L}^{2}(S^{1})$ onto itself. \textbf{B. }If we write $A$
in block matrix form with respect to the decomposition $\mathbf{L}^{2}%
(S^{1})=\mathbf{H}^{+}\oplus\mathbf{H}^{-}$ as
\begin{equation}
A=\left(
\begin{array}
[c]{cc}%
a & b\\
c & d
\end{array}
\right)  \label{block}%
\end{equation}
then, the operators $b$ and $c$ are Hilbert-Schmidt operators.
\end{definition}

In \cite{psegal} it is proved that $\mathbb{GL}_{\mathrm{res}}(\mathbf{L}%
^{2}(S^{1}))$ acts transitively on the Segal-Wilson Grassmannian so it is a
homogeneous space. It has a natural left invariant metric defined in
Section~7.8 of \cite{psegal}. It is enough to construct the metric at the
point $\mathbf{H}^{+}=\mathbf{H}^{+}\oplus\{0\}\in\mathbb{G}r_{\infty}$. Note
that
\[
T_{\mathbf{H}^{+},\mathbb{G}r_{\infty}}=\mathbb{HS}(\mathbf{H}^{+}%
\mathbf{,}\mathbf{H}^{-}),
\]
where $\mathbb{HS}(\mathbf{H}^{+}\mathbf{,H}^{-})$ denotes the space of
Hilbert-Schmidt operators from $\mathbf{H}^{+}$ into $\mathbf{H}^{-}$. Hence
there is a naturally defined scalar product on $T_{H_{+},\mathbb{G}r_{\infty}%
}$ by $\psi$ and $\chi$ in
\begin{equation}
\langle\psi,\chi\rangle\overset{def}{=}Tr(\psi^{\ast}\chi), \label{km}%
\end{equation}
where $\psi$ and $\chi\in T_{H_{+},\mathbb{G}r_{\infty}}$ \footnote{Our
definition differs from the one used in \cite{psegal} by a factor of 2}.

We proved in \cite{STZ} the following Theorem:

\begin{theorem}
\label{curv} Let $\left\{  \psi_{k}\right\}  $ be an orthonormal basis of
\[
T_{\mathbf{H}^{+},\mathbb{G}r_{\infty}}=\mathbb{HS}(\mathbf{H}^{+}%
\mathbf{,}\mathbf{H}^{-})\subset Hom\left(  \mathbf{H}^{+}\mathbf{,}%
\mathbf{H}^{-}\right)  .
\]
Then,
\end{theorem}

\textbf{a.} \textit{The left invariant metric defined by }$\left(
\ref{km}\right)  $\textit{ is K\"{a}hler and} \textit{the component of the
curvature tensor with respect to the orthonormal basis }$\left\{  \psi
_{k}\right\}  $\textit{ is given by:}%
\[
R_{i\overline{j},k\overline{l}}=-\delta_{i\overline{j}}\delta_{k\overline{l}%
}-\delta_{i\overline{l}}\delta_{k\overline{j}}+Tr\left(  \psi_{i}\psi
_{j}^{\ast}\wedge\psi_{k}\psi_{l}^{\ast}\right)  +Tr\left(  \psi_{i}\psi
_{l}^{\ast}\wedge\psi_{k}\psi_{j}^{\ast}\right)
\]
\textit{for }$\left(  i,j\right)  \neq\left(  k,l\right)  $\textit{ and
}$R_{i\overline{j},i\overline{j}}=-2\delta_{i\overline{j}}+Tr\left(  \psi
_{i}\psi_{j}^{\ast}\wedge\psi_{k}\psi_{l}^{\ast}\right)  .$ \textbf{b.
}\textit{Let }$\psi$\textit{ be any complex direction in the tangent space
}$T_{\mathbf{H}^{+}}\mathbb{G}r_{\infty}.$\textit{ Let }$K_{\psi}$\textit{ be
the Gaussian sectional curvature in any the two-dimensional real space defined
by }$\operatorname{Re}\psi$\textit{ and }$\operatorname{Im}\psi.$\textit{ Then
we have }$K\psi=-2+Tr\left(  \psi\psi^{\ast}\wedge\psi\psi^{\ast}\right)
<-\frac{3}{2}.$

\subsection{The Embedding of $\mathbb{H}^{-1,1}(\mathbb{D})$ into the Tangent
Space at a point of the Segal-Wilson Grassmannian}

\begin{theorem}
\label{Pr1}Let $\mu_{\psi}\in\mathbb{H}^{-1,1}(\mathbb{D}),$ where
$\mathbb{H}^{-1,1}(\mathbb{D})$ is defined by $\left(  \ref{2a}\right)  .$ Let
us define the operator $A_{\mu_{\psi}}:\mathbf{H}^{+}\rightarrow\mathbf{H}%
^{-}$ affiliated with the Beltrami differential $\mu_{\psi}\left(
\overline{dz}\otimes\frac{d}{dz}\right)  $ as follows:%
\begin{equation}
A_{\mu_{\psi}}(f(z)):=\frac{1}{\lambda}\mathbf{pr}_{\lambda\mu_{\psi}}%
^{-}\left(  \left(  \mathbf{pr}_{\lambda\mu_{\psi}}^{+}\right)  ^{-1}\left(
f(z)\right)  \right)  , \label{emb1}%
\end{equation}
where $\lambda$ is such a positive number that
\[
\left\Vert \lambda\mu_{\psi}\right\Vert _{\mathbf{L}^{\infty}(\mathbb{D}%
)}=\left\Vert \lambda(1-|z|^{2})^{2}\psi(\overline{z})\right\Vert
_{\mathbf{L}^{\infty}(\mathbb{D})}\leq k<1.
\]
Then t\textit{he map}
\begin{equation}
\iota:\mu_{\psi}\left(  \overline{dz}\otimes\frac{d}{dz}\right)  \rightarrow
A_{\mu_{\psi}}, \label{Emb}%
\end{equation}
\textit{defines an embedding of the Hilbert space }$\mathbb{H}^{-1,1}%
(\mathbb{D})$ \textit{into the Hilbert space} $T_{\mathbf{H}^{+}%
,\mathbb{G}r_{\infty}}=\mathbb{HS}\left(  \mathbf{H}^{+},\mathbf{H}%
^{-}\right)  $ and\textbf{ so }$\iota\left(  \mathbb{H}^{-1,1}(\mathbb{D}%
)\right)  $ \textit{is a closed Hilbert subspace in} $T_{\mathbf{H}%
^{+},\mathbb{G}r_{\infty}}=\mathbb{HS}\left(  \mathbf{H}^{+},\mathbf{H}%
^{-}\right)  .$
\end{theorem}

\textbf{Proof of part 1: }The definition of $\mathbb{H}^{-1,1}(\mathbb{D}),$
Theorem \ref{PrB} and Theorem \ref{Pr} imply that the map given by $\left(
\ref{Emb}\right)  $ defines a map from $\mathbb{H}^{-1,1}(\mathbb{D})$ to the
space of Hilbert-Schmidt operators $\mathbb{HS}\left(  H_{+},H_{-}\right)  .$
The proof of Theorem \ref{Pr1} follows from the Lemma bellow:

\begin{lemma}
\label{Pr1a}Let $\mu_{\psi_{1}}$ and $\mu_{\psi_{2}}$ be two different
elements of $\mathbb{H}_{1}^{-1,1}(\mathbb{D}).$ Then $A_{\mu_{\psi_{1}}}\neq
A_{\mu_{\psi_{2}}}$ in $\mathbb{HS}\left(  \mathbf{H}^{+},\mathbf{H}%
^{-}\right)  .$
\end{lemma}

\textbf{Proof: }The definition of the linear operators $A_{\mu_{\psi_{i}}}$
implies that $\mathbf{W}_{\mu_{\psi_{i}}}$ are the graphs of the operators
$A_{\mu_{\psi_{i}}}.$ If we prove that the graphs $\mathbf{W}_{\mu_{\psi_{1}}%
}$ and $\mathbf{W}_{\mu_{\psi_{2}}}$ of the operators $A_{\mu_{\psi_{i}}}$are
different Hilbert subspaces in $\mathbf{L}^{2}(S^{1})$ then $A_{\mu_{\psi_{1}%
}}\neq A_{\mu_{\psi_{2}}}.$

Let us consider the unique quasi-conformal maps $\Phi_{i}(z)$ of
$\mathbb{C}\cup\infty:=\mathbb{CP}^{1}$ defined by
\[
\left(  \overline{\partial}-\mu_{\psi_{i}}\partial\right)  \left(  \Phi
_{i}(z)\right)  =0
\]
for $i=1,2$ which satisfy condition $\left(  \ref{bel6}\right)  $ of Theorem
\ref{B4}$.$ Let $\Phi_{i,\infty}(z)=\Phi_{i}(z)\left\vert _{\mathbb{D}^{\ast}%
}\right.  .$ We will prove that the assumption $\psi_{1}(\overline{z})\neq
\psi_{2}(\overline{z})$ implies that $\Phi_{1,\infty}(\mathbb{D}^{\ast}%
)\neq\Phi_{2,\infty}(\mathbb{D}^{\ast}).$ Theorem \ref{Pr1} will follow.

\begin{proposition}
\label{Pr1c}Suppose that $\Phi_{1,\infty}(\mathbb{D}^{\ast})\neq\Phi
_{2,\infty}(\mathbb{D}^{\ast})$ then $\mathbf{W}_{\mu_{\psi_{1}}}%
\neq\mathbf{W}_{\mu_{\psi_{2}}}$ in $\mathbf{L}^{2}(S^{1}).$
\end{proposition}

\textbf{Proof: }If we normalize $\Phi_{i,\infty}(z)\frac{\sqrt{-1}}{2},%
{\displaystyle\int\limits_{\mathbb{D}^{\ast}}}
\left\vert \Phi_{i,\infty}(z)\right\vert ^{2}dz\wedge\overline{dz}=1,$ then
according to \cite{Ahlfors} the functions
\[
\left\{  \left(  \Phi_{i,\infty}(z)\right)  ^{n},\text{ for }n=1,...\right\}
\]
are holomorphic in $\Phi_{i,\infty}(\mathbb{D}^{\ast})$ and form an orthogonal
bases for $\mathbf{W}_{\mu_{\psi_{i}}}.$ So from this fact we deduce that
$\mathbf{W}_{\mu_{\psi_{1}}}=\mathbf{W}_{\mu_{\psi_{2}}}$ if an only if
$\Phi_{1,\infty}(\mathbb{D}^{\ast})=\Phi_{2,\infty}(\mathbb{D}^{\ast}).$
Proposition \ref{Pr1c} is proved. $\blacksquare$

\begin{proposition}
\label{Pr1d}$\psi_{1}(\overline{z})\neq\psi_{2}(\overline{z})$ implies
$\Phi_{1,\infty}(\mathbb{D}^{\ast})\neq\Phi_{2,\infty}(\mathbb{D}^{\ast}).$
\end{proposition}

\textbf{Proof: }The proof of Proposition \ref{Pr1d} follows from the following

We assumed that the functions $\psi_{1}(\overline{z})$ is different from
$\psi_{2}(\overline{z}).$ So the definition of $\mu_{\psi}(z)=\left(
1-|z|^{2}\right)  ^{2}\psi(\overline{z})$ implies $\mu_{\psi_{1}}\neq\mu
_{\psi_{2}}\Longleftrightarrow\psi_{1}\neq\psi_{2}.$

\begin{proposition}
\label{diff}Suppose that $\psi_{1}(\overline{z})\neq\psi_{2}(\overline{z})$
then for any conformal map $A$ of $\mathbb{D}^{\ast},$ $\Phi_{1,\infty
}(\mathbb{D}^{\ast})\neq\Phi_{2,\infty}\left(  A\left(  \mathbb{D}^{\ast
}\right)  \right)  .$
\end{proposition}

\textbf{Proof: }Suppose that $\psi_{1}(\overline{z})\neq\psi_{2}(\overline
{z})$ and for some conformal map $A$ of $\mathbb{D}^{\ast}$ we have
$\Phi_{1,\infty}(\mathbb{D}^{\ast})=\Phi_{2,\infty}\left(  A\left(
\mathbb{D}^{\ast}\right)  \right)  .$ We will show that this assumption
contradicts Theorem \ref{AW}. One of the basic properties of the Schwarzian
states
\[
\mathcal{S}\left[  \left.  \Phi\left(  A\left(  \frac{1}{z}\right)  \right)
\right\vert _{\mathbb{D}^{\ast}}\right]  =\mathcal{S}\left[  \left.
\Phi\left(  \frac{1}{z}\right)  \right\vert _{\mathbb{D}^{\ast}}\right]
\]
for any conformal $A$ map of $\mathbb{D}^{\ast}.$ Theorem \ref{AW} implies
that
\[
\psi_{1}\left(  z\right)  =\mathcal{S}\left[  \left.  \Phi_{1,\infty}\left(
A\left(  \frac{1}{z}\right)  \right)  \right\vert _{\mathbb{D}^{\ast}}\right]
=\psi_{2}\left(  z\right)  =\mathcal{S}\left[  \left.  \Phi_{2,\infty}\left(
A\left(  \frac{1}{z}\right)  \right)  \right\vert _{\mathbb{D}^{\ast}}\right]
.
\]
So we get a contradiction. Proposition \ref{diff} is proved. $\blacksquare$
Lemma \ref{Pr1a} is proved. $\blacksquare$

\begin{lemma}
\label{pair2}Recall that $\mathbb{H}_{1}^{-1,1}(\mathbb{D})$ is the subset in
$\mathbb{H}^{-1,1}(\mathbb{D})$ which consists of functions $(1-|z|^{2}%
)^{2}\psi(\overline{z}),$ where $\psi(\overline{z})$ is antiholomorphic
function in the unit disk,
\[
\left.  \psi(\overline{z})\right\vert _{S^{1}}\in\mathbf{H}_{S^{1},h}^{-3/2}%
\]
and
\[
\left\Vert (1-|z|^{2})^{2}\psi(\overline{z})\right\Vert _{\mathbf{L}^{\infty
}(\mathbb{D})}\leq k<1.
\]
Let $\iota_{3\left/  2\right.  }:\mathbb{H}_{1}^{-1,1}(\mathbb{D}%
)\rightarrow\mathbf{H}_{S^{1},h}^{3\left/  2\right.  }$ be a linear map
defined by:
\[
\iota_{3\left/  2\right.  }\left(  \left(  1-|z|^{2}\right)  ^{2}%
\psi(\overline{z})\right)  =\Psi\left(  \overline{z}\right)  \left\vert
_{S^{1}}\right.  ,
\]
where $\Psi\left(  \overline{z}\right)  $ is given by $\left(  \ref{3a}%
\right)  .$ Then $\iota_{3\left/  2\right.  }$ is a continuous map and
$\iota_{3\left/  2\right.  }\left(  \mathbb{H}_{1}^{-1,1}(\mathbb{D})\right)
$ is an open set in $\mathbf{H}_{S^{1},h}^{3\left/  2\right.  }.$
\end{lemma}

\textbf{Proof: }Let $\Psi_{0}(\overline{z})=\iota_{3\left/  2\right.  }\left(
\mathbb{(}1-|z|^{2}\psi(\overline{z})\right)  \in\mathbf{H}_{S^{1}%
,h}^{3\left/  2\right.  }.$ We need to show that there exists $\varepsilon
,\delta>0$ such that if%
\begin{equation}
\underset{|z|<1}{\sup}\left(  1-|z|^{2}\right)  ^{2}\left\vert \psi
(\overline{z})-\psi_{0}(\overline{z})\right\vert <\delta\label{A2}%
\end{equation}
then
\[
\left\Vert \Psi(\overline{z})-\Psi_{0}(\overline{z})\right\Vert _{\mathbf{H}%
_{S^{1},h}^{3\left/  2\right.  }}^{2}<\varepsilon.
\]
We proved that
\[
\left\Vert \Psi(\overline{z})-\Psi_{0}(\overline{z})\right\Vert _{\mathbf{H}%
_{S^{1},h}^{3\left/  2\right.  }}^{2}=\left\Vert \psi(\overline{z})-\phi
_{0}(\overline{z})\right\Vert _{\mathbf{H}_{S^{1},h}^{-3\left/  2\right.  }%
}^{2}.
\]
Recall that if
\[
\psi(\overline{z})=%
{\displaystyle\sum\limits_{n=2}^{\infty}}
a_{n}\overline{z}^{n-2},
\]
then
\[
\Psi(\overline{z})=%
{\displaystyle\sum\limits_{n=2}^{\infty}}
\frac{a_{n}}{n\left(  n^{2}-1\right)  }\overline{z}^{n-2}=%
{\displaystyle\sum\limits_{n=2}^{\infty}}
b_{n}\overline{z}^{n-2}.
\]
Then $\left(  \ref{A2}\right)  $ implies that
\[
\left\Vert \left(  \psi(\overline{z})-\psi_{0}(\overline{z})\right)
\left\vert _{S^{1}}\right.  \right\Vert _{\mathbf{H}_{S^{1},h}^{-3\left/
2\right.  }}^{2}=
\]%
\[
\frac{\sqrt{-1}}{2}%
{\displaystyle\int\limits_{\mathbb{D}}}
\left(  1-|z|^{2}\right)  ^{2}\left\vert \psi(\overline{z})-\psi_{0}%
(\overline{z})\right\vert dz\wedge\overline{dz}\leq
\]%
\[
\left(  \underset{|z|<1}{\sup}\left(  1-|z|^{2}\right)  ^{2}\left\vert
\psi(\overline{z})-\psi_{0}(\overline{z})\right\vert \right)  \frac{\sqrt{-1}%
}{2}%
{\displaystyle\int\limits_{\mathbb{D}}}
dz\wedge\overline{dz}<\frac{\delta}{2\pi}.
\]
On the other hand we have
\[
\left\Vert \left(  \psi(\overline{z})-\psi_{0}(\overline{z})\right)
\left\vert _{S^{1}}\right.  \right\Vert _{\mathbf{H}_{S^{1},h}^{-3\left/
2\right.  }}^{2}=\left\Vert \Psi(\overline{z})-\Psi_{0}(\overline
{z})\right\Vert _{\mathbf{H}_{S^{1},h}^{3\left/  2\right.  }}^{2}<\frac
{\delta}{2\pi}.
\]
Lemma \ref{pair2} is proved since $\varepsilon=\frac{\delta}{2\pi}$.
$\blacksquare$

According to Lemma \ref{pair2} $\mathbb{H}_{1}^{1,1}\left(  \mathbb{D}\right)
$ is an open set in $T_{id,\mathbf{T}^{\infty}}^{3/2}.$ Nag proved that the
map $\iota:\mathbb{H}_{1}^{1,1}\left(  \mathbb{D}\right)  \rightarrow
\mathbb{HS}\left(  \mathbf{H}^{+},\mathbf{H}^{-}\right)  $ is an isometric
embedding when restricted on all $\mu_{\psi}$ such that $\psi(\overline
{z})\left\vert _{S^{1}}\right.  \in C^{\infty}\left(  \mathbb{C}\right)  $
with respect to the left invariant metrics. According to Lemma \ref{pair2}
$\mathbb{H}_{1}^{1,1}\left(  \mathbb{D}\right)  $ is an open set in
$T_{id,\mathbf{T}^{\infty}}^{3/2}.$ Thus the linear map $\iota$ is an isometry
and one to one on an open and everywhere dense subset in $\mathbb{H}_{1}%
^{1,1}\left(  \mathbb{D}\right)  $ into the Hilbert space $\mathbb{HS}\left(
\mathbf{H}^{+},\mathbf{H}^{-}\right)  .$ This implies that $\iota$ is a
continuous map from the open set $\mathbb{H}_{1}^{1,1}\left(  \mathbb{D}%
\right)  $ of the Hilbert space in $\mathbb{H}^{1,1}\left(  \mathbb{D}\right)
$ into the Hilbert space $\mathbb{HS}\left(  \mathbf{H}^{+},\mathbf{H}%
^{-}\right)  .$ So $\iota$ will be a continuos linear map of the Hilbert space
$\mathbb{H}^{1,1}\left(  \mathbb{D}\right)  $ into the Hilbert space
$\mathbb{HS}\left(  \mathbf{H}^{+},\mathbf{H}^{-}\right)  .$ Theorem \ref{Pr1}
is proved. $\blacksquare$

\subsection{Hilbert $3\left/  2\right.  $ Manifold Structure on the
$\mathbf{T}^{\infty}$ and the Exponential Map}

\begin{definition}
We recall that a submanifold $S$ of a Riemannian manifold $M$ is called
totally geodesic at $p$ if each $M$-geodesic passing through $p$ in a tangent
direction to $S$ remains in $S$ for all time. If $S$ is geodesic at all its
points, then it is called totally geodesic~\cite{helgason}.
\end{definition}

Theorem \ref{G} was proved in \cite{STZ}. We reproduced the proof of Theorem
\ref{G} in the Appendix of the paper. 

\begin{theorem}
\label{G}Let $\psi\in\mathbb{HS}(\mathbf{H}^{+}\mathbf{,H}^{-})=T_{\mathbf{H}%
^{+},\mathbb{G}r_{\infty}}.$ Let $\exp$ be the map from $\mathbb{HS}%
(\mathbf{H}^{+}\mathbf{,H}^{-})=T_{H_{+},\mathbb{G}r_{\infty}}$ to
$\mathbb{G}r_{\infty}$ defined by the left invariant metric$.$ Then the
complex curve $\gamma_{\psi}(s)=\exp\left(  s\psi\right)  \subset
\mathbb{G}r_{\infty}$ is a totally geodesic complex submanifold of complex
dimension one in $\mathbb{G}r_{\infty}$ and $\gamma_{\psi}(s)=\exp\left(
s\psi\right)  $ exists for all $s\in\mathbb{C}.$
\end{theorem}

The following analogue of the infinite dimensional complex analytic analogue
of Hadamard's theorem was proved in \cite{STZ} based on Theorem \ref{G}:

\begin{theorem}
\label{exp}The complex exponential map defined in Theorem \ref{G} is a
covering complex analytic one to one map from the tangent space
$T_{id,\mathbb{G}r_{\infty}}$ onto $\mathbb{G}r_{\infty}$. We reproduced the
proof of Theorem \ref{exp} in the Appendix of the paper. 
\end{theorem}

Based on Theorems \ref{G} and \ref{exp} by using very easy and standard
arguments we will give a very simple proof of the following Theorem:

\begin{theorem}
\label{3/2}The completion $\mathbf{T}^{3\left/  2\right.  }$ of $\mathbf{T}%
^{\infty}$ with respect to the left invariant K\"{a}hler metric defined by
$\left(  \ref{3/2pair}\right)  $\textbf{ }is complex analytic Hilbert manifold
isomorphic to the totally geodesic closed Hilbert submanifold $\exp\left(
\iota\left(  \mathbb{H}^{-1,1}(\mathbb{D})\right)  \right)  $ in
$\mathbb{G}r_{\infty},$ where $\iota:\mathbb{H}^{-1,1}(\mathbb{D}%
)\rightarrow\mathbb{HS}\left(  \mathbf{H}^{+}\mathbf{,H}^{-}\right)  $ is the
embedding defined in Theorem \ref{Pr1}.
\end{theorem}

\textbf{Proof: }A Theorem of Nag proved in \cite{nag} that $\mathbf{T}%
^{\infty}$ with the left invariant K\"{a}hler metric is isometrically embedded
$\mathbb{G}r_{\infty}.$ For each complex direction
\[
\psi\in\left(  \mathbf{H}_{S^{1},h}^{3\left/  2\right.  }\right)
\approxeq\iota\left(  \mathbb{H}^{-1,1}(\mathbb{D})\right)  \subset
T_{\mathbf{H}^{+},\mathbb{G}r_{\infty}}=\mathbb{HS}(\mathbf{H}^{+}%
,\mathbf{H}^{-})
\]
the exponential map defines a complex analytic isomorphism between the complex
line $s\psi$ and totally geodesic one dimensional complex submanifold
\[
D_{\psi}:=\gamma(s)=\exp\left(  s\psi\right)  \subset\mathbb{G}r_{\infty}%
\]
in the direction $\psi.$\ According to Theorem \ref{Pr1} $\iota\left(
\mathbb{H}^{-1,1}(\mathbb{D})\right)  $ is a closed Hilbert subspace in
$T_{\mathbf{H}_{+},\mathbb{G}r_{\infty}}.$ Thus Theorem \ref{cov} implies that
$\mathbf{T}^{3\left/  2\right.  }:=\exp\left(  \iota\left(  \mathbb{H}%
^{-1,1}(\mathbb{D})\right)  \right)  $ is a closed totally geodesic complex
analytic submanifold in $\mathbb{G}r_{\infty}$ containing the image of
$\mathbf{T}^{\infty}$ in $\mathbb{G}r_{\infty}.$ According to Cor. \ref{CS3/2}
the Hilbert space $\mathbb{H}^{-1,1}(\mathbb{D})$ is isomorphic to the Hilbert
space $\mathbf{H}_{S^{1},h}^{3/2}.$ Theorem \ref{3/2} is proved.
$\blacksquare$

\begin{corollary}
\label{3/2a}The complex analytic Hilbert structure on $\mathbf{T}^{3\left/
2\right.  }$ is modeled by the Sobolev $3/2$ space $\mathbf{H}_{S^{1},h}%
^{3/2}.$
\end{corollary}

\textbf{Proof: }Corollary follows directly from Theorem \ref{Co} and Theorem
\ref{3/2}. $\blacksquare$

\begin{definition}
By definition the infinite-dimensional Siegel disc is the set of
Hilbert-Schmidt operators $T:\mathbf{H}^{+}\mathbf{\rightarrow H}^{-}$ are
such that $\det(I-TT^{\ast})>0$.
\end{definition}

\begin{corollary}
\label{exp2} $\mathbf{T}^{3/2}$ $\approxeq\exp\left(  \iota\left(
\mathbb{H}^{-1,1}(\mathbb{D})\right)  \right)  $ is a totally geodesic complex
analytic Hilbert submanifold isomorphic to the infinite-dimensional Siegel
disc defined by the Hilbert-Schmidt operators $T\in\iota\left(  \mathbb{H}%
^{-1,1}(\mathbb{D})\right)  \subseteq\mathbb{HS}\left(  \mathbf{H}%
^{+},\mathbf{H}^{-}\right)  $.
\end{corollary}

\textbf{Proof: }Theorems \ref{G}, \ref{exp} and the definition of the complex
analytic exponential map imply that $\exp\left(  \iota\left(  \mathbb{H}%
^{-1,1}(\mathbb{D})\right)  \right)  $ is a totally geodesic closed complex
analytic Hilbert submanifold in $\mathbb{G}r_{\infty}.$ We proved that any two
points in both spaces can be joined by a unique geodesic and $\exp\left(
\iota\left(  \mathbb{H}^{-1,1}(\mathbb{D})\right)  \right)  $ is geodesically
complete. This implies that $\exp\left(  \iota\left(  \mathbb{H}%
^{-1,1}(\mathbb{D})\right)  \right)  $ can be isometrically identified with
the infinite-dimensional Siegel disc defined by the Hilbert-Schmidt operators
$T\in\iota\left(  \mathbb{H}^{-1,1}(\mathbb{D})\right)  .$ $\blacksquare$

\begin{corollary}
\label{CKM}The space $\mathbf{T}^{3\left/  2\right.  }:=\left(  \mathbf{Diff}%
_{+}^{\infty}(S^{1})/\mathbb{PSU}_{1,1}\right)  ^{3\left/  2\right.  }$
equipped with the unique invariant K\"{a}hler metric has negative curvature in
holomorphic directions. More precisely, the curvature is negative and
uniformly bounded away from zero in holomorphic directions.
\end{corollary}

\textbf{Proof: }According to the Theorem of Nag proved in \cite{nag}, the
embedding of $\mathbf{T}^{\infty}$ into the Grassmannian is isometric. Since
$\mathbf{T}^{\infty}$ is an everywhere dense subset in $\mathbf{T}^{3/2},$
Theorem \ref{Pr1} implies the Hilbert manifold $\mathbf{T}^{3\left/  2\right.
}$ is isometrically embedded into Segal-Wilson Grassmannian $\mathbb{G}%
r_{\infty}.$ Cor. \ref{CKM} followed directly from Theorem \ref{curv}.
$\blacksquare$

\section{Appendix}

\subsection{Differential Geometry of the Segel Wilson Grassmannian}

\textbf{Theorem }\ref{curv}. \textit{Let }$\left\{  \psi_{k}\right\}
$\textit{ be an orthonormal basis of }%
\[
T_{\mathbf{H}^{+}}\mathbb{G}r_{\infty}=\mathbb{HS}(\mathbf{H}^{+}%
\mathbf{,}\mathbf{H}^{-})\subset Hom\left(  \mathbf{H}^{+}\mathbf{,}%
\mathbf{H}^{-}\right)  .
\]
\textit{Then,}

\textbf{a.} \textit{The left invariant metric defined by }$\left(
\ref{km}\right)  $\textit{ is K\"{a}hler and} \textit{the component of the
curvature tensor with respect to the orthonormal basis }$\left\{  \psi
_{k}\right\}  $\textit{ is given by:}%
\[
R_{i\overline{j},k\overline{l}}=-\delta_{i\overline{j}}\delta_{k\overline{l}%
}-\delta_{i\overline{l}}\delta_{k\overline{j}}+Tr\left(  \psi_{i}\psi
_{j}^{\ast}\wedge\psi_{k}\psi_{l}^{\ast}\right)  +Tr\left(  \psi_{i}\psi
_{l}^{\ast}\wedge\psi_{k}\psi_{j}^{\ast}\right)
\]
\textit{for }$\left(  i,j\right)  \neq\left(  k,l\right)  $\textit{ and
}$R_{i\overline{j},i\overline{j}}=-2\delta_{i\overline{j}}+Tr\left(  \psi
_{i}\psi_{j}^{\ast}\wedge\psi_{k}\psi_{l}^{\ast}\right)  .$ \textbf{b.
}\textit{Let }$\psi$\textit{ be any complex direction in the tangent space
}$T_{\mathbf{H}^{+}}\mathbb{G}r_{\infty}.$\textit{ Let }$K_{\psi}$\textit{ be
the Gaussian sectional curvature in any the two-dimensional real space defined
by }$\operatorname{Re}\psi$\textit{ and }$\operatorname{Im}\psi.$\textit{ Then
we have }%
\[
K\psi=-2+Tr\left(  \psi\psi^{\ast}\wedge\psi\psi^{\ast}\right)  <-\frac{3}%
{2}.
\]

\textbf{Proof of part a}: The proof is based on the construction of the
so-called Cartan coordinate system in \cite{LTYZ}. By that we mean a
holomorphic coordinate system $(x^{1},x^{2},\dots)$ in which the components of
the K\"{a}hler metric tensor $g_{i\overline{j}}(x)$ is given by the formula:
$g_{i\overline{j}}=\delta_{i\overline{j}}+r_{i\overline{j},i\overline{j}}%
x^{k}\bar{x}^{l}+\mathcal{O}(|x|^{3}).$ Cartan proved that if the coordinate
system satisfies this last equation, then $r_{i\overline{j},i\overline{j}%
}=-\frac{1}{3}R_{i\overline{j},i\overline{j}}.$ where $R_{i\overline
{j},k\overline{l}}$ is the curvature tensor. To prove part \textbf{a)} we
construct the Cartan coordinates in $\mathbb{G}r_{\infty}$:

\textbf{Definition A.48.1 }\textit{Let }$\psi_{i}$ \textit{be an orthonormal
basis of }$\mathbb{HS}(\mathbf{H}^{+}\mathbf{,H}^{-}).$ \textit{Let}
\[
\varphi_{t}=%
{\displaystyle\sum\limits_{i=1}^{\infty}}
t_{i}\psi_{i}\in\mathbb{HS}(\mathbf{H}^{+}\mathbf{,}\mathbf{H}^{-}).
\]
\textit{Let the subspace }$\mathbf{W}_{t}$ \textit{of }$\mathbf{H}^{+}%
\oplus\mathbf{H}^{-}$ \textit{be spanned by the set}
\[
\{1+\varphi_{t}(1),z+\varphi_{t}(z),\dots,z^{n}+\varphi_{t}(z^{n}),\dots\}.
\]
\textit{Obviously, }$\mathbf{W}_{t}$ \textit{is the graph of the operator
}$\varphi_{t}$. \textit{We define the exponential map}
\[
\exp:\mathbb{HS}(\mathbf{H}^{+}\mathbf{,}\mathbf{H}^{-})\rightarrow
\mathbb{G}r_{\infty}%
\]
\textit{as follows}
\begin{equation}
\exp\left(
{\displaystyle\sum\limits_{i=1}^{\infty}}
t_{i}\psi_{i}\right)  =\mathbf{W}_{t}\in\mathbb{G}r_{\infty}.\label{Exp}%
\end{equation}
\textit{Thus }$t=(t_{1},t_{2},\dots)$\textit{ will define the local
coordinates in some open set }$U\subset\mathbb{G}r_{\infty}$ \textit{of the
point }$0$ \textit{corresponding to} $\mathbf{H}^{+}\in\mathbb{G}r_{\infty}$.

\textbf{Lemma A.48.2 }\textit{The following expansion near }$t=0$\textit{ in
the coordinates }$t=(t_{1},t_{2},\dots)$\textit{ holds}%
\[
-\frac{\partial^{2}}{\partial t^{i}\overline{\partial t^{j}}}\log\det\left(
1-%
{\displaystyle\sum\limits_{i,j}}
t^{i}\overline{t^{j}}\psi_{i}\psi_{j}^{\ast}\right)  =
\]%
\[
\delta_{i\overline{j}}+\left(  2\delta_{i\overline{j}}-Tr\left(  \psi_{i}%
\psi_{j}^{\ast}\wedge\psi_{j}\psi_{i}^{\ast}\right)  \right)  t^{i}%
\overline{t^{j}}+
\]%
\[%
{\displaystyle\sum\limits_{\left(  i,j\right)  \neq(k,l)}}
\left(  \delta_{i\overline{j}}\delta_{k\overline{l}}+\delta_{i\overline{l}%
}\delta_{k\overline{j}}-Tr\left(  \psi_{i}\psi_{j}^{\ast}\wedge\psi_{k}%
\psi_{l}^{\ast}\right)  -Tr\left(  \psi_{i}\psi_{l}^{\ast}\wedge\psi_{k}%
\psi_{j}^{\ast}\right)  \right)  t^{k}\overline{t^{l}}.
\]

\textbf{Proof:} Let $f(t)=\det\left(  id-\sum t^{i}\overline{t^{j}}\psi
_{i}\psi_{j}^{\ast}\right)  ,$ then
\begin{equation}
\frac{\partial^{2}}{\partial t^{i}\overline{\partial t^{j}}}\log\det\left(
id-\sum t^{i}\overline{t^{j}}\psi_{i}\psi_{j}^{\ast}\right)  =\frac
{\partial^{2}}{\partial t^{i}\overline{\partial t^{j}}}f^{-1}-\frac{\partial
f}{\partial t^{i}}\frac{\overline{\partial}f}{\overline{\partial}%
\overline{t^{j}}}f^{-2}.\label{eq:eq1}%
\end{equation}
From the definition of the determinant, we have that%
\[
\det\left(  id-\sum_{i,j}t^{i}\overline{t^{j}}\psi_{i}\psi_{j}^{\ast}\right)
=
\]%
\[
1-\sum_{i,j}\mathrm{Tr}(\psi_{i}\psi_{j}^{\ast})t^{i}\overline{t^{j}}%
+\sum_{i,j,k,l}\mathrm{Tr}(\psi_{i}\psi_{j}^{\ast}\wedge\psi_{k}\psi_{l}%
^{\ast})t^{i}\overline{t^{j}}t^{k}\overline{t^{l}}+\mathrm{h.o.t.}%
\]
Hence,%
\[
\frac{\partial^{2}}{\partial t^{i}\overline{\partial t^{j}}}\det\left(
id-\sum t^{i}\overline{t^{j}}\psi_{i}\psi_{j}^{\ast}\right)  =-\delta
_{i\bar{j}}+\mathrm{Tr}(\psi_{i}\psi_{j}^{\ast}\wedge\psi_{i}\psi_{j}^{\ast
}))t^{i}\overline{t^{j}}+
\]%
\[
\sum_{(k,l)\neq(i,j)}\mathrm{Tr}\left(  \psi_{i}\psi_{j}^{\ast}\wedge\psi
_{k}\psi_{l}^{\ast}+\psi_{i}\psi_{l}^{\ast}\wedge\psi_{k}\psi_{j}^{\ast
}\right)  t^{k}\overline{t^{l}}+\mathrm{h.o.t},
\]%
\[
\frac{\partial}{\partial t^{i}}\det\left(  id-\sum_{i,j}t^{i}\overline{t^{j}%
}\psi_{i}\psi_{j}^{\ast}\right)  =-\sum_{l}\mathrm{Tr}(\psi_{i}\psi_{l}^{\ast
})\bar{t}_{l}+\mathrm{h.o.t.},
\]
and
\[
\frac{\overline{\partial}}{\overline{\partial}\overline{t^{j}}}\det\left(
id-\sum t^{i}\overline{t^{j}}\psi_{i}\psi_{j}^{\ast}\right)  =-\sum
_{k}\mathrm{Tr}(\psi_{k}\psi_{j}^{\ast})t^{k}+\mathrm{h.o.t.}.
\]
Furthermore,
\[
\frac{1}{\det\left(  id-\sum t^{i}\overline{t^{j}}\psi_{i}\psi_{j}^{\ast
}\right)  }=1+\sum t^{i}\overline{t^{j}}\mathrm{Tr}(\psi_{i}\psi_{j}^{\ast
})+\mathrm{h.o.t}%
\]
and
\[
1/f^{2}=1+2\sum t^{i}\overline{t^{j}}\mathrm{Tr}(\psi_{i}\psi_{j}^{\ast
})+\mathrm{h.o.t.}%
\]
Substituting the last 5 equations into equation (\ref{eq:eq1}) we obtain the
result. Lemma \textbf{A.48.2} is proved. $\blacksquare$

\textbf{Lemma} \textbf{A.48.2} implies that the coordinates $(t_{1}%
,t_{2},\dots)$ forms a Cartan coordinate system. Thus, for $(i,j)\neq(k,l)$ we
get:
\[
R_{i\overline{j},k\overline{l}}=-\delta_{i\overline{j}}\delta_{k\overline{l}%
}-\delta_{i\overline{l}}\delta_{k\overline{j}}+\mathrm{Tr}(\psi_{i}\psi
_{j}^{\ast}\wedge\psi_{k}\psi_{l}^{\ast}+\psi_{i}\psi_{l}^{\ast}\wedge\psi
_{k}\psi_{j}^{\ast}),
\]
and
\[
R_{i\overline{j},i\overline{j}}=-2\delta_{i\overline{j}}+\mathrm{Tr}(\psi
_{i}\psi_{j}^{\ast}\wedge\psi_{i}\psi_{j}^{\ast}).
\]
So we proved

\textbf{Corollary A.48.3 }\textit{The left invariant metric on} $\mathbb{G}%
r_{\infty}$ \textit{is K\"{a}hler with a potential }$\log\det\left(  id-\sum
t^{i}\overline{t^{j}}\psi_{i}\psi_{j}^{\ast}\right)  $.

This concludes the proof of \textbf{part~a} of Theorem \ref{curv}.
$\blacksquare$.

\textbf{Proof of part b: }To prove part~b) remark that the Gaussian curvature
in the direction $\psi_{i}$ is given by
\[
K_{\psi_{i}}=R_{i\overline{j},i\overline{j}}=-2+\mathrm{Tr}(\psi_{i}\psi
_{i}^{\ast}\wedge\psi_{i}\psi_{i}^{\ast}).
\]
Now, if $\psi_{1}=\varphi$, we complete the set $\{\psi_{1}\}$ to an
orthonormal set in the Hilbert space $T_{H_{+},\mathbb{G}r_{\infty}}$. Hence,
from the previous discussion it follows that $K_{\varphi}=-2+\mathrm{Tr}%
(\varphi\varphi^{\ast}\wedge\varphi\varphi^{\ast}).$

\textbf{Lemma A.48.4 }\textit{If }$Tr\left(  \psi\psi^{\ast}\right)
=1,$\textit{ then }$Tr\left(  \wedge^{2}\psi\psi^{\ast}\right)  <\frac{1}{2}.$

\textbf{Proof: }Note that $\varphi\varphi^{\ast}$ is a compact positive
operator and hence its nonzero eigenvalues are all positive. Let
$\{\lambda_{i}\}$ be the set of nonzero eigenvalues of $\varphi\varphi^{\ast}%
$. Since $\varphi\varphi^{\ast}$ is trace-class and $||\varphi||^{2}%
=\mathrm{Tr}\left(  \varphi\varphi^{\ast}\right)  =1$ imply that
\[
\mathrm{Tr}\left(  \varphi\varphi^{\ast}\right)  =\sum_{i}\lambda_{i}=1.
\]
A simple argument with tensor products gives
\[
Tr(\wedge^{2}\left(  \varphi\varphi^{\ast}\right)  )=%
{\displaystyle\sum\limits_{i<j}}
\lambda_{i}\lambda_{j}.
\]
We have
\[
1=\left(
{\displaystyle\sum\limits_{i}}
\lambda_{i}\right)  ^{2}=2%
{\displaystyle\sum\limits_{i<j}}
\lambda_{i}\lambda_{j}+%
{\displaystyle\sum\limits_{i}}
\lambda_{i}^{2}.
\]
From here it follows that $Tr(\wedge^{2}\varphi\varphi^{\ast})<1/2.$ Using the
formula for the curvature in holomorphic direction $\varphi$ it follows that
$K_{\varphi}<-3\left/  2\right.  <0.$ $\blacksquare$ \textbf{Lemma A.48.4}
implies part \textbf{2} of Theorem \ref{curv}. Theorem \ref{curv} is proved.
$\blacksquare$

\textbf{Definition A.48.5 }\textit{We recall that a submanifold }$S$\textit{
of a Riemannian manifold }M\textit{ is called totally geodesic at }$p$\textit{
if each }M\textit{-geodesic passing through }$p$\textit{ in a tangent
direction to }$S$\textit{ remains in }$S$\textit{ for all time. If }%
$S$\textit{ is geodesic at all its points, then it is called totally
geodesic~}\cite{helgason}.

Theorem \ref{G} was proved in \cite{STZ}. We reproduce the proof. The reason
is to make this paper self contained.

\textbf{Theorem \ref{G}. }\textit{Let} $\psi\in\mathbb{HS}(\mathbf{H}%
^{+}\mathbf{,H}^{-})=T_{\mathbf{H}^{+},\mathbb{G}r_{\infty}}.$ \textit{Let}
$\exp$ \textit{be the map from} $\mathbb{HS}(\mathbf{H}^{+}\mathbf{,H}%
^{-})=T_{H_{+},\mathbb{G}r_{\infty}}$ \textit{to} $\mathbb{G}r_{\infty}$
\textit{defined by} \textbf{Definition A.48.1} \textit{Then the complex curve
}$\gamma_{\psi}(s)=\exp\left(  s\psi\right)  \subset\mathbb{G}r_{\infty}$
\textit{is a totally geodesic complex submanifold of complex dimension one in}
$\mathbb{G}r_{\infty}$ \textit{and} $\gamma_{\psi}(s)=\exp\left(
s\psi\right)  $ \textit{exists for all }$s\in\mathbb{C}.$

\textbf{Proof: }The proof of the Theorem \ref{G} is a consequence of the
following two results:

\textbf{Lemma A.56.1 }\textit{Let} $\psi\in\mathbb{HS}\left(  \mathbf{H}%
^{+}\mathbf{,H}^{-}\right)  $ \textit{be such that} $\left\Vert \psi
\right\Vert _{HS}^{2}=1$ \textit{and}
\[
s(t)=s_{0}+te^{i\theta},
\]
\textit{with }$\theta,t\in\mathbb{R}$ \textit{be a line in} $\mathbb{R}$.
\textit{Let }$\gamma_{\psi}(s(t)):=\exp\left(  s(t)\psi\right)  $ \textit{be a
curve in} $\mathbb{G}r_{\infty}.$ \textit{Then, the norm of the tangent
vector} $\dot{\gamma}_{\psi}$ \textit{to the the path }$\gamma_{\psi}(s(t))$
\textit{has a length one, i.e. }$\left\Vert \dot{\gamma}_{\psi}\left(
s(t)\right)  \right\Vert =1.$

\textbf{Proof:} By the definition of the exponential map the line $s(t)\psi$
in the Hilbert space $\mathbb{HS}\left(  \mathbf{H}^{+}\mathbf{,H}^{-}\right)
$ corresponds to the family of Hilbert subspaces $\mathbf{W}_{s}:=\gamma
_{\psi}(s)\subset\mathbf{H}^{+}\mathbf{\oplus H}^{-}$ in $\mathbb{G}r_{\infty
}$ spanned by the vectors
\[
s(t)\psi(1),...,s(t)\psi(z^{n}),....
\]
With respect to the decomposition $\mathbf{H}=\mathbf{H}^{+}\oplus
\mathbf{H}^{-}$ define the block matrix
\[
A_{s}=\left(
\begin{array}
[c]{cc}%
id & \overline{s}\psi^{\ast}\\
s\psi & id
\end{array}
\right)
\]
From the definition of $\mathbf{W}_{s}$\textbf{ }we have $\mathbf{W}_{s}%
=A_{s}\mathbf{H}^{+}.$ Let
\[
T:\mathbf{H}^{+}\oplus\mathbf{H}^{-}\mathbf{\rightarrow}\mathbf{H}^{+}%
\oplus\mathbf{H}^{-}%
\]
be any bounded operator. Using that $Graph(T)$ is perpendicular to
$Graph^{\prime}\left(  T^{\ast}\right)  $, where
\[
Graph^{\prime}\left(  T\right)  \overset{def}{=}\{(-Tx,x)\ \left\vert x\in
Dom(T)\right.  \ \},
\]
we get $\mathbf{W}_{s}^{\perp}=A_{s}\mathbf{H}^{-}.$ Hence, the matrix $A_{s}$
maps $\mathbf{H=H}^{+}\mathbf{\oplus H}^{-}$ into $\mathbf{W}_{s}%
\mathbf{\oplus W}_{s}^{\perp}$ and preserves the direct sum decomposition. It
is well known that the operators $(id+s\bar{s}\psi\psi^{\ast})$ and
$(id+s\bar{s}\psi^{\ast}\psi)$ are invertible. The following relations can be
checked easily:
\begin{equation}
(id+s\overline{s}\psi^{\ast}\psi)^{-1}\psi^{\ast}=\psi^{\ast}(id+s\overline
{s}\psi\psi^{\ast})^{-1},\label{eqa1}%
\end{equation}
and
\begin{equation}
(id+s\overline{s}\psi\psi^{\ast})^{-1}\psi=\psi(id+s\overline{s}\psi^{\ast
}\psi)^{-1}.\label{eqa2}%
\end{equation}
Therefore,%
\[
A_{s}^{-1}=\left(
\begin{array}
[c]{cc}%
id & \overline{s}\psi^{\ast}\\
s\psi & id
\end{array}
\right)  \left(
\begin{array}
[c]{cc}%
(id+s\overline{s}\psi^{\ast}\psi)^{-1} & 0\\
0 & (id+s\overline{s}\psi\psi^{\ast})^{-1}%
\end{array}
\right)  =
\]%
\begin{equation}
\left(
\begin{array}
[c]{cc}%
(id+s\overline{s}\psi^{\ast}\psi)^{-1} & 0\\
0 & (id+s\overline{s}\psi\psi^{\ast})^{-1}%
\end{array}
\right)  \left(
\begin{array}
[c]{cc}%
id & \overline{s}\psi^{\ast}\\
s\psi & id
\end{array}
\right)  .\label{eqa3}%
\end{equation}
To compute the norm $||\dot{\gamma}_{\psi}(s(t))||$, we use the identification
of $T_{\mathcal{W}_{s(t)},\mathbb{G}r_{\infty}}$ with $\mathbb{HS}%
(\mathbf{W}_{s(t)},\mathbf{W}_{s(t)}^{\perp})$. The latter, is mapped onto
$\mathbb{HS}(\mathbf{H}^{+}\mathbf{,H}^{-})$ by means of $\widetilde{X}\mapsto
A_{s}^{-1}\widetilde{X}A_{s}\left\vert _{\mathbf{H}^{+}}\right.  .$ From the
invariance of the metric we have that the norm of$\ \widetilde{X}\in
T_{\mathbf{W},\mathbb{G}r_{\infty}}$ is given by
\begin{equation}
||\widetilde{X}||^{2}=\mathrm{Tr}\left(  (A_{s}^{-1}\widetilde{X}%
A_{s}\left\vert _{\mathbf{H}^{+}}\right.  )^{\ast}(A_{s}^{-1}\widetilde
{X}A_{s}\left\vert _{\mathbf{H}^{+}}\right.  .)\right)  .\label{vareq1}%
\end{equation}
Since,
\[
\dot{\gamma}_{\psi}(s(t))=\left(
\begin{array}
[c]{cc}%
0 & -e^{-i\theta}\psi^{\ast}\\
e^{i\theta}\psi & 0
\end{array}
\right)  ,
\]
a technical but straightforward computation with $\widetilde{X}=\dot{\gamma
}_{\psi}(s(t))$ taking into account equations~(\ref{eqa1}), (\ref{eqa2}), and
(\ref{eqa3}) yields the Lemma \textbf{A.56.1}. $\blacksquare$\hfill

\textbf{Lemma A.56.2} \textit{Set} $\dot{\gamma}_{\psi}(s)=\frac{d}{ds}%
\gamma_{\psi}(s)$ \textit{and assume that }$\left\Vert \dot{\gamma}_{\psi
}(s)\right\Vert =1.$ \textit{Let }$\nabla$ \textit{be the covariant derivative
in }$\mathbb{G}r_{\infty}$ \textit{given by the Levi-Civita connection. Then,}%
\[
\nabla_{\dot{\gamma}_{\psi}(s)}\left(  \dot{\gamma}_{\psi}(s)\right)  =0.
\]

\textbf{Proof:} Let $s\in\mathbb{R}$ and $\gamma_{\psi}(s)$ be a geodesic in
our K\"{a}hler manifold. For each point $s$ of the geodesic $\gamma_{\psi}(s)$
we define a complex direction as follows:
\[
\dot{\gamma_{\psi}}(s)+i\mathcal{I}\dot{\gamma_{\psi}}(s).
\]
For each $s$ in $\gamma_{\psi}(s)$ we consider a geodesic $\mathcal{I}%
\dot{\gamma_{\psi}}(s)$ and so each point $\tau$ on the geodesic from $s$ with
direction $\mathcal{I}\dot{\gamma_{\psi}}(s)$ we have two tangent vectors. The
first, $\alpha(\tau)$, which is the parallel transport of $\dot{\gamma_{\psi}%
}(s)$ and the second $\beta(\tau)$ which is given by
\[
\beta(\tau)=\frac{d}{ds}\mathcal{I}\dot{\gamma_{\psi}}(s),
\]
and is the parallel transport of $\mathcal{I}\dot{\gamma_{\psi}}$. The
K\"{a}hler condition implies that the complex structure operator is parallel
with respect to the Levi-Civita connection of the metric. From here we get
that for the Lie bracket $[\dot{\gamma}(s),\mathcal{I}\dot{\gamma}(s)]$ of the
vector fields we have $[\dot{\gamma}(s),\mathcal{I}\dot{\gamma}(s)]=0.$ Hence,
it follows from Frobenius theorem that there exist a surface $S$ such that the
tangent space is spanned by $\alpha(s)$ and $\beta(s)$. Since,
\[
\mathcal{I}(\alpha(s)+i\beta(s))=i(\alpha(s)+i\beta(s))
\]
it follow that $S$ is a complex analytic curve and we can take $z$ as a
complex analytic coordinate associated to the point
\[
z=\exp(x\dot{\gamma_{\psi}}(s)+iy\mathcal{I}\dot{\gamma_{\psi}}(s)).
\]
Let's write $\dot{\mu}(z)=\dot{\gamma_{\psi}}(s)+i\mathcal{I}\dot{\gamma
_{\psi}}(s).$ From the properties of the Levi-Civita connection and
$\left\Vert \dot{\mu}(z)\right\Vert ^{2}=const$ it follows that
\begin{equation}
0=\frac{d}{dz}||\dot{\mu}(z)||^{2}\left\vert _{s_{0}}\right.  =2\langle
\nabla_{z}\dot{\mu}(z),\dot{\mu}(z)\rangle=0,\label{eq:eq0}%
\end{equation}
and so
\[
\frac{d^{2}}{d\bar{z}dz}||\dot{\mu}(z)||^{2}\left\vert _{s_{0}}\right.
=||\nabla_{z}\dot{\mu}(z)||^{2}-R_{\psi}||\dot{\mu}(z)||^{2}=0.
\]
Since $R_{\psi}<0$, we have $\nabla_{z}\dot{\mu}(z)=0.$ Restrict $\nabla
_{z}\dot{\mu}(z)$ on the real part at $s_{0}$ and we get that $\nabla_{z}%
\dot{\gamma_{\psi}}(z)=0.$ Hence, taking $\mathrm{Re}z=\dot{\gamma_{\psi}}$,
we have $\nabla_{\dot{\gamma_{\psi}}(s)}\dot{\gamma_{\psi}}(s)\left\vert
_{s_{0}}\right.  =0.$ Lemma \textbf{A.56.2}  is proved. $\blacksquare$ Theorem
\ref{G} is proved. $\blacksquare$

In this section we will prove the infinite dimensional complex analytic
analogue of Hadamard's theorem, to show the existence of the complex analytic
manifold structure $\mathbb{H}^{-1,1}(\mathbb{D})$ on $\mathbf{T}^{\infty}$.

\textbf{Theorem \ref{exp}. }\textit{The complex exponential map defined in
Theorem \ref{G} is a covering complex analytic one to one map from the tangent
space} $T_{id,\mathbb{G}r_{\infty}}$ \textit{onto} $\mathbb{G}r_{\infty}$.

\textbf{Proof: }The proof Theorem \ref{exp} is based on Theorem \ref{G}.

\textbf{Lemma A.57.1} \textit{Let} $\psi\in T_{\mathbf{H}^{+},\mathbb{G}%
r_{\infty}}=\mathbb{HS}\left(  \mathbf{H}^{+}\mathbf{,H}^{-}\right)  $
\textit{and} $\left\Vert \psi\right\Vert _{HS}=1.$ \textit{Then the
exponential map restricted to one dimensional complex subspace}
\[
\{c\psi|c\in{\mathbb{{C}}},\text{ }\psi\in\mathbb{HS}\left(  \mathbf{H}%
^{+}\mathbf{,}\mathbf{H}^{-}\right)  \}\subset T_{\mathbf{H}^{+}%
,\mathbb{G}r_{\infty}}=\mathbb{HS}\left(  \mathbf{H}^{+}\mathbf{,}%
\mathbf{H}^{-}\right)
\]
\textit{and taking values in the totally geodesic submanifold }$D_{\psi}%
$\textit{ is a complex analytic diffeomorphisms.}

\textbf{Proof:} To prove Lemma \textbf{A.57.1} we use that the curvature in
the holomorphic direction is negative. This fact implies that we have
\begin{equation}
||\left(  d\exp(c\overrightarrow{w})\right)  \left(  \overrightarrow
{w}\right)  ||\geq||\overrightarrow{w}||.\label{ext}%
\end{equation}
For the proof see \cite{helgason}. From a standard result in (finite
dimensional) differential geometry states that $\left(  \ref{ext}\right)  $
implies that the map is a local diffeomorphisms and thus it is covering.
Furthermore, if $c_{1}\neq c_{2}$ then
\[
\mathbf{W}_{c_{1}\psi}:=\left(  id+c_{1}\psi\right)  \left(  H_{+}\right)
\neq\mathbf{W}_{c_{2}\psi}:=\left(  id+c_{2}\psi\right)  \left(
\mathbf{H}^{+}\right)  .
\]
Hence, $\exp(c_{1}\psi)\neq\exp(c_{2}\psi).$ Thus the exponential map
\[
\exp:c\psi\rightarrow\exp\left(  c\psi\right)  =D_{\psi}%
\]
is a diffeomorphisms. Next we will show that it is complex analytic map. We
need to show that if $\overrightarrow{v}\in T_{\exp(c_{1}\psi),D_{\psi}},$
then $J\left(  \overrightarrow{v}\right)  \in T_{\exp(c_{1}\psi),D_{\psi}},$
where $J$ is the complex structure operator$.$ $\overrightarrow{v}$ is
obtained by a parallel transform of a vector $\overrightarrow{v}_{0}\in
T_{0,D_{\psi}}$ along the total geodesic submanifold $D_{\psi}.$ We know that
$J\left(  \overrightarrow{v}_{0}\right)  \in T_{0,D_{\psi}}.$ Since the metric
is K\"{a}hler then $\nabla J=0.$ Since $D_{\psi}$ is a totally geodesic
submanifold, then by the parallel transposition $\overrightarrow{v}_{0}\in
T_{0,D_{\psi}}$ goes to $\overrightarrow{v}\in T_{\exp(c_{1}\psi),D_{\psi}}.$
So $\nabla J=0$ and $D_{\psi}$ is a totally geodesic submanifold imply that by
parallel transport of $J\left(  \overrightarrow{v}_{0}\right)  \in
T_{0,D_{\psi}}$ goes to $J\left(  \overrightarrow{v}\right)  \in T_{\exp
(c_{1}\psi),D_{\psi}}.$ Lemma \textbf{A.57.1} is proved. $\blacksquare$

Suppose that $\psi_{1}$ and $\psi_{2}$ are linearly independent vectors. Then
the construction of $\exp$ and $\left(  \ref{ext}\right)  $ imply that complex
curves $\{\exp(t\psi_{1})|t\in{\mathbb{{C}}}\}$ and $\{\exp(t\psi_{2}%
)|t\in{\mathbb{{C}}}\}$ intersect only at the identity. These arguments imply
that $\exp$ is a complex analytic isomorphism of some open set
$T_{id,\mathbb{G}r_{\infty}}$ to an open set in $\mathbb{G}r_{\infty}$. Since
the curvature of the K\"{a}hler metric is negative in complex direction then
$\left(  \ref{ext}\right)  $ implies that the exponential map is surjective.
Theorem \ref{cov} is proved. $\blacksquare$

\end{document}